\documentclass[10pt]{article}

\usepackage{amsmath}
\usepackage{graphicx}
\usepackage{indentfirst}
\usepackage{subfigure}
\usepackage{cite}
\usepackage{color}

\begin{document}

\title{\bf{Dilaton Field Released under Collision of Dilatonic Black Holes with Gauss-Bonnet Term}}

\date{}
\maketitle

\begin{center}
\author{Bogeun Gwak}$^a$\footnote{rasenis@sejong.ac.kr} and \author{Daeho Ro}$^b$\footnote{daeho.ro@apctp.org }\\

\vskip 0.25in
$^{a}$\it{Department of Physics and Astronomy, Sejong University, Seoul 05006, Republic of Korea}\\
$^{b}$\it{Asia Pacific Center for Theoretical Physics, POSTECH, Pohang, Gyeongbuk 37673, Republic of Korea}\\
\end{center}
\vskip 0.6in

{\abstract
{
We investigate the upper limit of the gravitational radiation released upon the collision of two dilatonic black holes by analyzing the Gauss-Bonnet term. Dilatonic black holes have a dilaton hair coupled with this term. Using the laws of thermodynamics, the upper limit of the radiation is obtained, which reflected the effects of the dilaton hair. The amount of radiation released is greater than that emitted by a Schwarzschild black hole due to the contribution from the dilaton hair. In the collision, most of the dilaton hair can be released through radiation, where the energy radiated by the dilaton hair is maximized when the horizon of one black hole is minimized for a fixed second black hole.
}}

\thispagestyle{empty}
\newpage
\setcounter{page}{1}

\section{Introduction}

Gravitational waves have been detected by the Laser Interferometer Gravitational-Wave Observatory (LIGO)\cite{Abbott:2016blz,Abbott:2016nmj,Abbott:2017vtc}. The sources of the waves have been the mergers of binary black holes in which the masses of the black holes have been more than 10 times the mass of the sun. The binary system that caused GW150914 consisted of black holes with masses of approximately $36M_\odot$ and $29M_\odot$ in the source frame\cite{Abbott:2016blz}. The recently detected gravitational wave, GW151226, was generated by a binary black hole merger involving two black holes with masses of $14.2M_\odot$ and $7.5M_\odot$\cite{Abbott:2016nmj}. The detections of these waves have proven that there are many black holes in our universe and that collisions between them may be frequent events.

For an asymptotic observer, a black hole in the Einstein-Maxwell theory can be distinguished by its conserved quantities: mass, angular momentum, and electric charge\cite{Ruffini:1971bza,Bekenstein:1995un,Mayo:1996mv}. This concept is known as the no-hair theorem, in which charges cannot be observed outside of the event horizon of a black hole. In the theory in which gravity is coupled with Maxwell and antisymmetric tensor fields, the dilaton hair concept was first introduced in association with string theory\cite{Gibbons:1982ih,Gibbons:1987ps,Garfinkle:1990qj}. Since then, many kinds of hairs have been described in different gravity theories, such as those involving Maxwell and Yang-Mills fields\cite{Droz:1991cx,Lee:1991vy,Breitenlohner:1991aa,Lavrelashvili:1992cp,Torii:1993vm,Breitenlohner:1994di,Kleihaus:2011tg,Kleihaus:2014lba,Kleihaus:2016rgf}. One of them is dilaton gravity theory, which includes the Gauss-Bonnet term\cite{Boulware:1986dr,Callan:1988hs,Mignemi:1992pm,Campbell:1991kz,Campbell:1992hc}, a curvature-squared term given in the effective field theory of a heterotic string theory\cite{Callan:1985ia,Zwiebach:1985uq,Gross:1986mw} and topological in four dimensions, so that the equations of motion are the same as in Einstein gravity when the dilaton field is turned off\cite{Boulware:1986dr,Callan:1985ia}. In dilaton gravity theory, the black hole solution has a dilaton field outside the black hole horizon\cite{Kanti:1995vq,Torii:1996yi,Kanti:1997br,Torii:1998gm,Kokkotas:2015uma}, and thus, dilaton hairs are an exception to the no-hair theorem. Because dilaton hairs originate from the masses of black holes, dilaton hairs are secondary hairs that grow from the primary hairs of black holes\cite{Coleman:1991ku,Coleman:1991jf}. The presence of a dilaton hair changes the physical properties of the corresponding black hole, such as its stability and thermodynamics, which have been studied in various black holes coupled with dilaton fields and Gauss-Bonnet terms\cite{Horne:1992zy,Mann:1992yv,Lavrelashvili:1992ia,Gibbons:1994vm,Cai:1997ii,Cai:2001dz,Cai:2003gr,Kim:2007iw,Goldstein:2009cv,Cai:2013qga}.

As a counterexample to the no-hair theorem, a dilatonic black hole should be stable in our universe. The stability of a dilatonic black hole can be tested and identified based on the specific range in which the mass lies. The solution for a dilatonic black hole is convergent with that for a Schwarzschild black hole in the large mass limit. Thus, its stability is also similar to that of a Schwarzschild black hole in the same range. When the mass is low, its behavior is different. A dilatonic black hole becomes unstable below a certain mass known as the critical mass. Thus, a dilatonic black hole must possess a certain minimum mass to be stable. In addition, at its critical mass, the black hole possesses minimum entropy and thus can be related to the cosmological remnant\cite{Kanti:1997br,Torii:1998gm,Moura:2006pz,Moura:2012fq}. On the other hand, the solution includes a naked singularity that is not allowed under the cosmic censorship conjecture\cite{Penrose:1969pc}. The cosmic censorship conjecture for black holes prevents naked singularities, so black holes should have horizons. Kerr black holes were first investigated with reference to the abovementioned conjecture by the inclusion of a particle\cite{Wald1974548}, and many black holes have subsequently been studied in a similar manner\cite{Jacobson:2009kt,Saa:2011wq,Gao:2012ca,Barausse:2010ka,Barausse:2011vx,Colleoni:2015afa,Colleoni:2015ena,Hubeny:1998ga,Isoyama:2011ea,Aniceto:2015klq,Hod:2016hqx,Horowitz:2016ezu,Toth:2015cda,Rocha:2011wp,Gwak:2012hq,Gwak:2015sua,BouhmadiLopez:2010vc,Doukas:2010be,Lehner:2010pn,Gwak:2011rp, Figueras:2015hkb,Gwak:2015fsa,Gwak:2016gwj}.

Classically, a black hole cannot emit particles, so its mass is nondecreasing. However, a black hole can radiate particles via the quantum mechanical effect, and its temperature can be defined in terms of the emitted radiation. The Hawking temperature is proportional to the surface gravity of a black hole\cite{Hawking:1974sw,Hawking:1976de}. The horizon areas of black holes can only increase via physical processes, which is similar to the behavior of entropy in a thermal system. Using this similarity, the entropy of a black hole, called Bekenstein-Hawking entropy, is defined as being proportional to its horizon area\cite{Bekenstein:1973ur,Bekenstein:1974ax}. Then, a black hole can be defined as a thermal system in terms of its Hawking temperature and Bekenstein-Hawking entropy. The nonperturbed stability can be tested based on the thermodynamics of the black hole and can be described by a heat capacity. However, a dilatonic black hole is thermally unstable, as its heat capacity is negative, but at the same time, its Hawking temperature has a finite value\cite{Torii:1998gm}, which is similar to that of a Schwarzschild black hole. One of other tests for nonperturbed stability is the fragmentation instability of black holes. The fragmentation instability is based on the entropy preference, so a black hole near an extremal bound decays into fragmented black holes that are thermally stable and have greater entropy than a single black hole system. For example, a Myers-Perry (MP) black hole is defined in higher dimensions, and its angular momentum has no upper bound over five dimensions\cite{Myers:1986un}. Then, an MP black hole becomes unstable when the large angular momentum is sufficiently large due to its centrifugal force. Thermally, the entropy of one extremal MP black hole is less than that of fragmented MP black holes, so an MP black hole breaks into multiple MP black holes\cite{Emparan:2003sy}. The fragmentation instability gives similar results for perturbation\cite{Shibata:2009ad,Dias:2009iu,Dias:2010eu,Dias:2010maa,Durkee:2010qu,Murata:2012ct,Dias:2014eua}. This kind of instability can also be obtained in rotating or charged anti-de Sitter (AdS) black holes\cite{Gwak:2014xra,Gwak:2015ysa}. A dilatonic black hole with a Gauss-Bonnet term also has a complicated phase diagram related to fragmentation instability\cite{Ahn:2014fwa}.

The gravitational radiation released when two black holes collide can be described thermodynamically. The sum of the entropies of the separate black holes in the initial state should be less than the entropy of the final black hole after the collision\cite{Hawking:1971tu}. Using the second law of thermodynamics, the minimum mass of the final black hole can be obtained based on the initial conditions. Thus, the difference between the initial and final masses is the mass released in the form of gravitational radiation. For Kerr black holes, the gravitational radiation depends on the alignments of their rotation axes\cite{Schiff:1960gi,Wilkins:1970wap,Mashhoon:1971nm,Wald:1972sz}. The dependency also exists for MP black hole collisions\cite{Gwak:2016cbq}. Many types of interaction energy can be released in the form of radiation upon collision. One of these types of interaction energy is that of the spin interaction between the black holes. If one of the initial black holes is infinitesimally small, the potential energy of the spin interaction is identical to the radiation energy obtained using thermodynamics in Kerr\cite{Wald:1972sz} and Kerr-AdS black holes\cite{Gwak:2016icd}. More precise analysis can be conducted using numerical methods in relativity\cite{Smarr:1976qy,Smarr:1977fy,Smarr:1977uf}. In this case, the waveform of the gravitational radiation can be investigated for different initial conditions\cite{Witek:2010xi,Bantilan:2014sra,Bednarek:2015dga,Hirotani:2015fxp,Sperhake:2015siy,Barkett:2015wia,Hinderer:2016eia,Konoplya:2016pmh}.

In this study, we investigated the upper limit of the gravitational radiation released due to the collision of two dilatonic black holes through the Gauss-Bonnet term. During the collision process, the energy of the black hole system will be released as radiation. Most of the radiation energy originates from the mass of the system, and the rest comes from various interactions between the black holes. There are many interactions, such as angular momentum and Maxwell charge interactions, that can contribute to the emitted radiation. The dilaton field is also one means through which black holes can interact. To an asymptotic observer, the dilaton charge, which is a secondary hair, is included in the mass of the black hole, so it acts as a mass distribution similar to a dust distribution around the black hole. Although the dilaton field is not observed in our universe, information about the behaviors of dust-like mass distributions in black hole collisions can be obtained. However, in Einstein gravity, a black hole cannot be coupled with a scalar field due to the no-hair theorem. Thus, the extent to which a scalar field can contribute to radiation is not well studied. For this reason, using a black hole solution coupled with a dilaton field through the Gauss-Bonnet term, the contribution of the dilaton field to the radiation can be determined. There are differences between dilatonic and Schwarzschild black holes, based on which the radiation of a dilatonic black hole can be distinguished from that of a Schwarzschild black hole. Therefore, we will show in this report that the dilaton field sufficiently affects the gravitational radiation released due to the collision of two black holes coupled with a dilaton hair through the Gauss-Bonnet term.

This paper is organized as follows. In section~\ref{sec2}, we review the concept of dilatonic black holes, which can be numerically obtained from the equations of motion in Einstein gravity coupled with a dilaton field through the Gauss-Bonnet term. In addition, the behaviors of dilatonic black hole for given parameters are introduced. In section~\ref{sec3}, we demonstrate how the upper limit of the gravitational radiation that is thermally allowed can be obtained and employ it to illustrate the differences between dilatonic black holes and black holes in Einstein gravity. In particular, we consider the contribution of the dilaton hair, since the limit is clearly different and distinguishable from that of a Schwarzschild black hole. We also discuss our results along with those of the LIGO experiment. In section~\ref{sec5}, we briefly summarize our results.

\section{Dilatonic Black Holes with Gauss-Bonnet Term}\label{sec2}
A dilatonic black hole is a four-dimensional solution to the Einstein dilaton theory with the Gauss-Bonnet term given by\cite{Kanti:1995vq,Torii:1996yi,Kanti:1997br,Torii:1998gm}. The dilaton field is coupled with the Gauss-Bonnet term in the Lagrangian
\begin{equation} \label{eq:lagrangian}
{\cal L} = \dfrac{R}{2} - \dfrac{1}{2} \nabla_\mu \phi \nabla^\mu \phi + f(\phi) R^2_{{GB}}\,,
\end{equation}
where the spacetime curvature and dilaton field are denoted as $R$ and $\phi$, respectively. The Einstein constant $\kappa = 8\pi G$ is set equal to unity for simplicity. The Gauss-Bonnet term is $R^2_{{GB}} = R^2 - 4R_{\mu\nu}R^{\mu\nu} + R_{\mu\nu\rho\sigma}R^{\mu\nu\rho\sigma}$ and is coupled with a function of the dilaton field, $f(\phi) = \alpha e^{\gamma \phi}$. The dilaton field is a secondary hair whose source is the mass of the conserved charge of the black hole. The dilaton hair appears as an element coupled with the Gauss-Bonnet term. The dilaton field equation and Einstein equations can be obtained from Eq.~\eqref{eq:lagrangian} and are as follows:
\begin{eqnarray}
0 &=& \dfrac{1}{\sqrt{-g}}\partial_\mu \left( \sqrt{-g}\partial^\mu \phi \right) + f'(\phi) R^2_{{GB}}, \\
0 &=& R_{\mu\nu} - \dfrac{1}{2}g_{\mu\nu} R - \partial_\mu \phi \partial^\mu \phi + \dfrac{1}{2} g_{\mu\nu} \partial_\rho \phi \partial^\rho \phi + T^{{GB}}_{\mu\nu}\,,
\end{eqnarray}
in which the GB term contributes to the energy-momentum tensor $T^{{GB}}_{\mu\nu}$\cite{Nojiri:2005vv}. Then,
\begin{align} \label{eq:tgb}
T^{{GB}}_{\mu\nu} =& - 4 (\nabla_\mu \nabla_\nu f(\phi)) R + 4 g_{\mu\nu} (\nabla^2 f(\phi)) R + 8 (\nabla_\rho \nabla_\mu f(\phi)) R_\nu{}^\rho + 8 (\nabla_\rho \nabla_\nu f(\phi)) R_\mu{}^\rho \nonumber
\\
&- 8 (\nabla^2 f(\phi)) R_{\mu\nu} - 8 g_{\mu\nu} (\nabla_\rho \nabla_\sigma f(\phi)) R^{\rho\sigma} + 8 (\nabla^\rho \nabla^\sigma f(\phi)) R_{\mu\rho\nu\sigma},
\end{align}
where only the nonminimally coupled terms in four-dimensional spacetime are presented in \cite{DeWitt:1964}.

A dilatonic black hole is a spherically symmetric and asymptotically flat solution for which the ansatz is given as\cite{Kanti:1995vq,Torii:1996yi,Kanti:1997br,Torii:1998gm}
\begin{equation} \label{eq:metric}
ds^2 = - e^{X(r)} dt^2 + e^{Y(r)} dr^2 + r^2 (d\theta^2 + \sin^2 \theta d\varphi^2)\,,
\end{equation}
where the metric exponents $X$ and $Y$ only depend on the radial coordinate $r$. Then, the dilaton field equation is
\begin{equation} \label{eq:dilaton}
\phi '' + \phi' \left( \dfrac{X' - Y'}{2} + \dfrac{2}{r} \right) - \dfrac{4\alpha \gamma e^{\gamma \phi}}{r^2} \left( X' Y' e^{-Y} + (1-e^{-Y}) \left(X'' + \dfrac{X'}{2}(X' - Y') \right) \right)=0\,,
\end{equation}
and the $(tt)$, $(rr)$, and $(\theta\theta)$ components of Einstein's equations are
\begin{align} \label{eq:ee1}
&\dfrac{r {\phi'}^2}{2} + \dfrac{1 - e^{Y}}{r} - Y' \left(1 + \dfrac{4 \alpha \gamma e^{\gamma \phi} \phi'}{r} (1 - 3e^{-Y}) \right) + \dfrac{8 \alpha \gamma e^{\gamma \phi}}{r} (\phi'' + \gamma \phi'{}^2)(1 - e^{-Y})=0\,, \\ \label{eq:ee2}
&\dfrac{r {\phi'}^2}{2} - \dfrac{1 - e^{Y}}{r} - X' \left(1 + \dfrac{4 \alpha \gamma e^{\gamma \phi} \phi'}{r} (1 - 3e^{-Y}) \right)=0\,, \\ \label{eq:ee3}
& X'' + \left(\dfrac{X'}{2} + \dfrac{1}{r}\right)(X'-Y') + \phi'{}^2 - \dfrac{8 \alpha \gamma e^{\gamma \phi - Y}}{r} \left(\phi' X'' + (\phi'' + \gamma \phi'{}^2)X' + \dfrac{\phi' X'}{2}(X'-3Y')\right)=0\,.
\end{align}
By taking the derivative of Eq.~\eqref{eq:ee2} with respect to $r$, $Y'$ can be eliminated from the equations of motion. The remaining equations of motion can be written as ordinary coupled differential equations:
\begin{equation} \label{eq:ode}
\phi'' = \dfrac{N_1}{D} \quad \text{and} \quad X'' = \dfrac{N_2}{D}\,,
\end{equation}
where $N_1$, $N_2$, and $D$ are only functions of $X'$, $Y$, $\phi$, and $\phi'$. The detailed expressions for these functions are given in Appendix \ref{sec:A}. The Gauss-Bonnet term is topological term in four-dimensional spacetime, so it cannot affect the equations of motion without the term $f(\phi)$ of the dilaton field. This idea can be easily shown by setting to $\phi=0$, where the dilaton field is turned off, so that $f(\phi)=\alpha$. However, the Gauss-Bonnet term still exists in Eq.~(\ref{eq:lagrangian}). Then, the dilaton field equation vanishes, and the equations of motion from Eqs.~(\ref{eq:ee1}) to (\ref{eq:ee3}) are reduced to
\begin{align}\label{eq:einstein1}
\frac{1-e^Y}{r}-Y'=0\,,\quad -\frac{1-e^Y}{r}-X'=0\,,\quad X''+\left(\frac{X'}{2}+\frac{1}{r}\right)(X'-Y')=0\,,
\end{align}
which are equations of motion for Einstein's gravity, $G_{\mu\nu}=0$. Hence, without the dilaton field, the effect of the Gauss-Bonnet term vanishes from the equations of motion.

\begin{figure}[h]
\centering\subfigure[{\small $\phi_h$ varies from $0.49483$ to $0.47368$ in $\alpha=1/16$ and $\gamma=\sqrt{2}$.}]
{\includegraphics[scale=0.52,keepaspectratio]{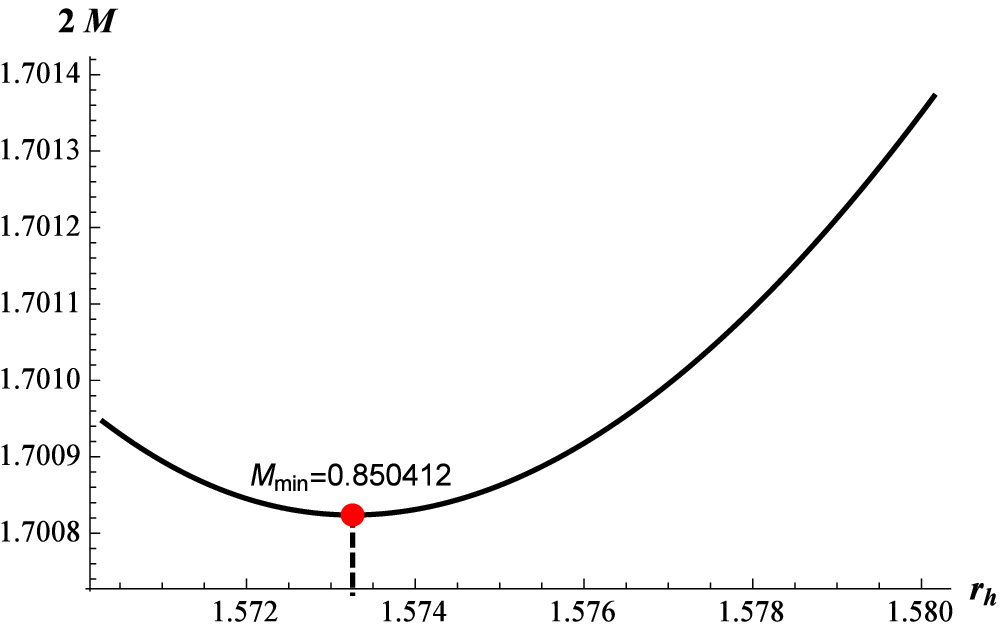}}
\quad
\centering\subfigure[{\small $\phi_h$ varies from $0.49829$ to $0.47791$ in $\alpha=1/16$ and $\gamma=1.35$.}]
{\includegraphics[scale=0.52,keepaspectratio]{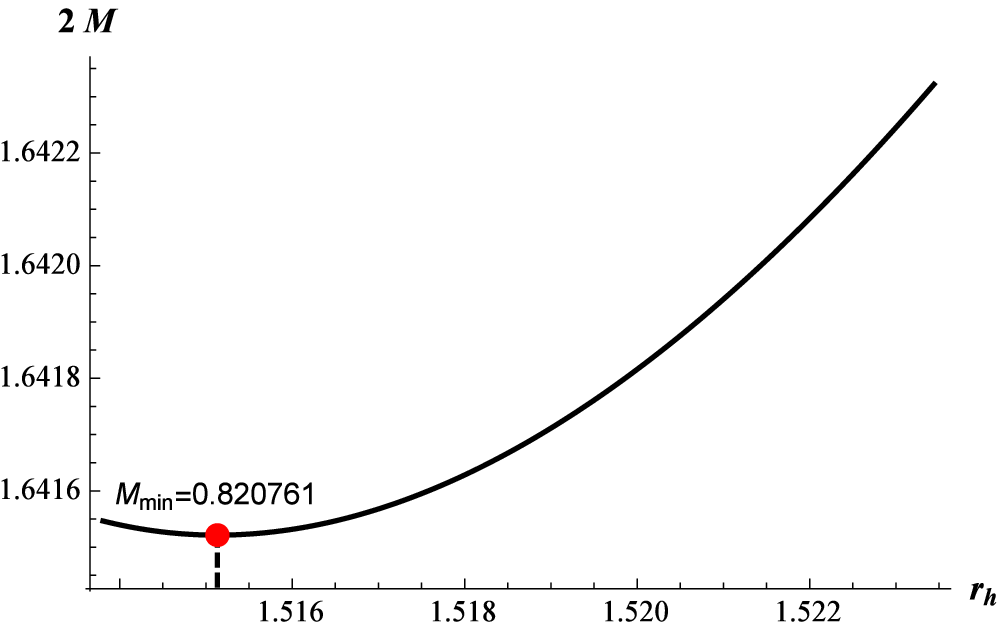}}
\quad
\centering\subfigure[{\small $\phi_h$ varies from $0.50141$ to $0.48137$ in $\alpha=1/16$ and $\gamma=1.29859$.}]
{\includegraphics[scale=0.52,keepaspectratio]{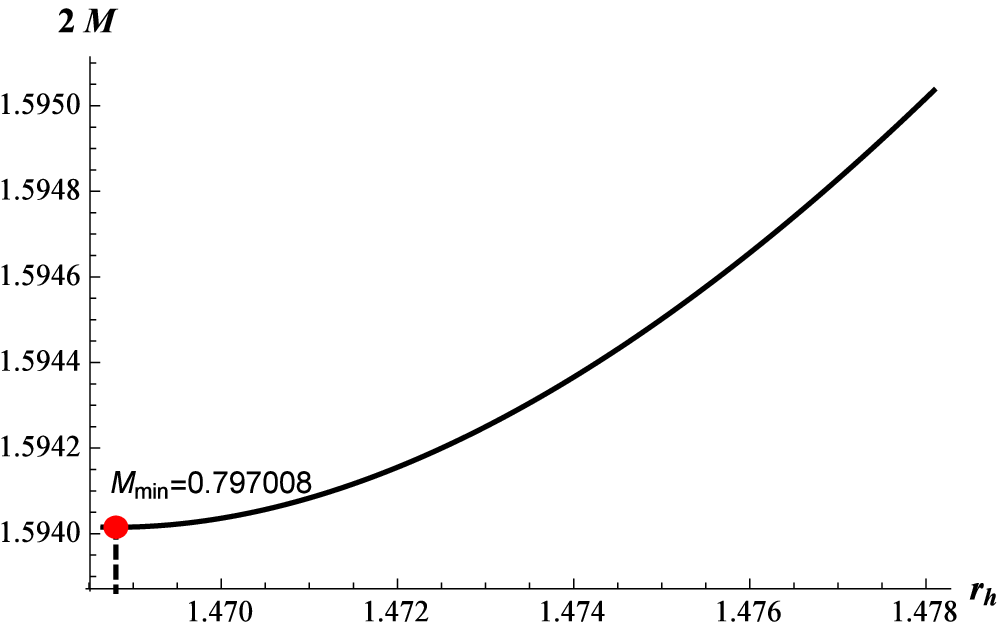}}
\caption{{\small The horizon versus mass of the dilatonic black hole with different $\gamma$.}}
\label{fig:solgamma}
\end{figure}

\begin{figure}[h]
\centering\subfigure[{\small $\alpha=1/16$ and $\gamma=\sqrt{2}$.}]
{\includegraphics[scale=0.80,keepaspectratio]{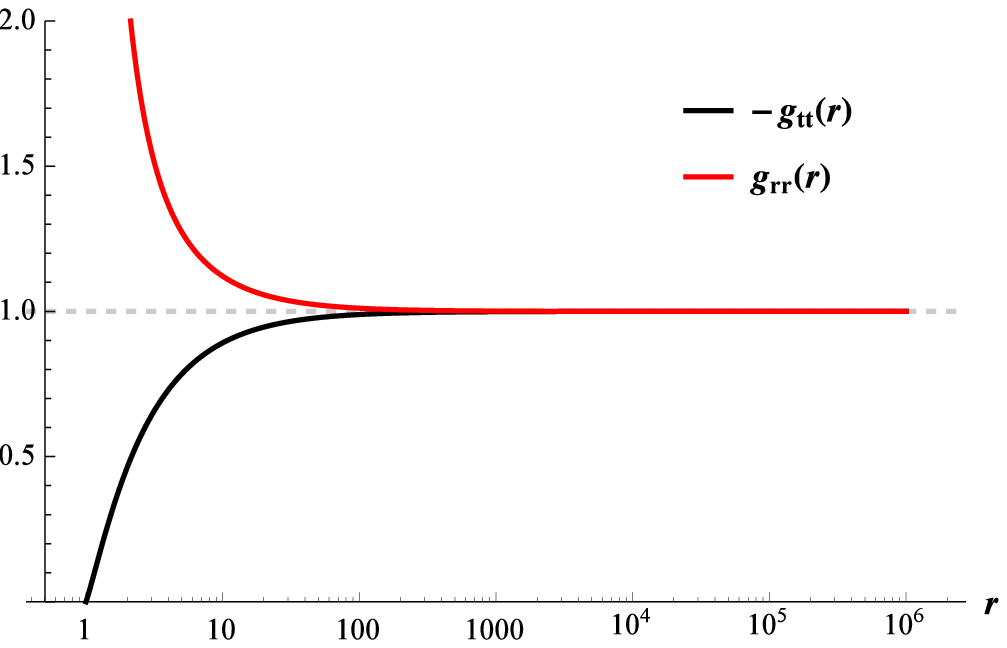}}
\quad
\centering\subfigure[{{\small $\alpha=1/16$ and $\gamma=\sqrt{2}$.}}]
{\includegraphics[scale=0.80,keepaspectratio]{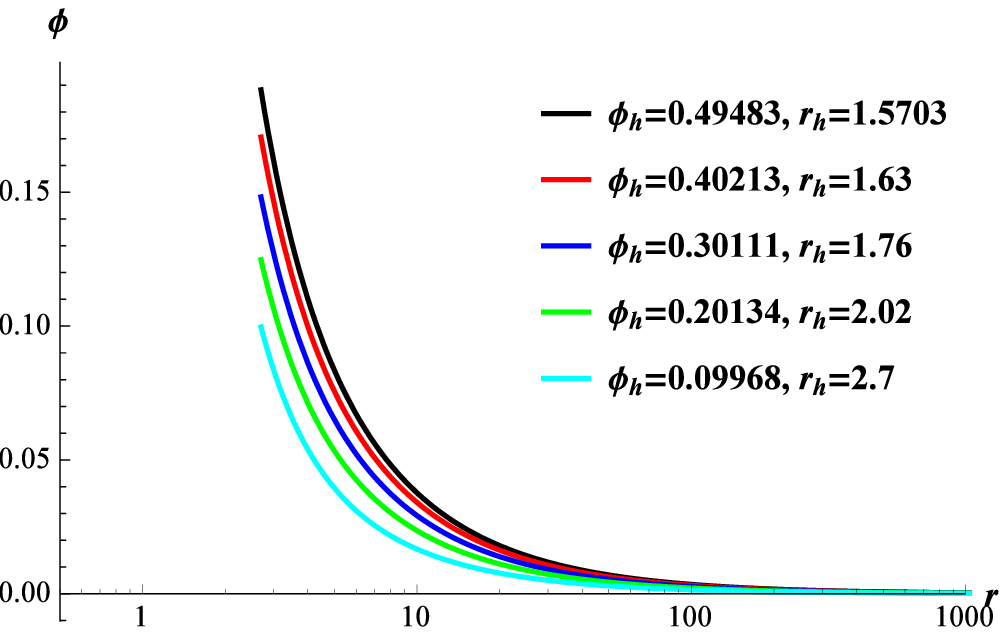}}
\caption{{\small The metric components and $\phi(r)$ of the dilatonic black holes.}}
\label{fig:solgamma2}
\end{figure}

The solution for a dilatonic black hole can be obtained by numerically solving Eq.~(\ref{eq:ode}). The numerical solution will be found from the outer horizon to infinity, so an initial condition at the outer horizon where the coordinate singularity is located is required. To determine the initial condition for the differential equations, it is necessary to investigate the behavior of a dilatonic black hole in the near-horizon region $r_h$. For the corresponding parameters at the horizon, the subscript $h$ is used. At the outer horizon, the metric should satisfy the relation $g_{rr}(r_h) = \infty$ or $g^{rr} = 0$. The metric components can be expanded in the near-horizon limit as
\begin{eqnarray} \label{eq:exhe}
e^{-X(r)} &=& x_1(r-r_h) + x_2 (r-r_h)^2 + \cdots, \\ \label{eq:eyhe}
e^{Y(r)} &=& y_1(r-r_h) + y_2 (r-r_h)^2 + \cdots, \\ \label{eq:phihe}
\phi(r) &=& \phi_h + \phi'_h(r-r_h) + \phi''_h (r-r_h)^2 + \cdots.
\end{eqnarray}
To check the divergence of $e^{Y(r)}$ at the outer horizon, $e^{Y}$ can be obtained using Eq.~\eqref{eq:ee2}:
\begin{equation} \label{eq:ey}
e^{Y(r)} = \dfrac{1}{4} \left(\big(2 - r^2 \phi'{}^2 + (2r + 8 \alpha \gamma e^{\gamma \phi} \phi') X'\big) + \sqrt{\big(2 - r^2 \phi'^2 + (2r + 8 \alpha \gamma e^{\gamma \phi} \phi') X'\big)^2 - 192 \alpha \gamma e^{\gamma \phi} \phi' X'} \right)\,.
\end{equation}
where the positive root has been chosen to form the horizon. From Eq.~(\ref{eq:ey}), $e^{Y(r)}$ has the same divergence of $X'(r)$. Furthermore, the plus sign was chosen in Eq.~(\ref{eq:ey}) to obtain the positive definition of $e^{Y(r)}$ in the limit of $X'$ going to the infinity at the outer horizon. The initial value of $e^{Y(r_h)}$ can be determined based on the values of other fields, such as $X'(r_h)$, $\phi_h$, and $\phi'_h$. To obtain a general solution, it is necessary to assume that $\phi_h$ and $\phi'_h$ are finite and that $X'$ tends to infinity when $r$ approaches the horizon as a result of Eq.~\eqref{eq:ey}. By considering series expansion up to $1/X'$ near the horizon, Eq.~\eqref{eq:ey} becomes
\begin{equation} \label{eq:eyh}
e^{Y(r)} = (r + 4 \alpha \gamma e^{\gamma \phi} \phi') X' + \dfrac{2r - r^3 \phi'{}^2 - 16 \alpha \gamma e^{\gamma \phi} \phi' - 4 r^2 \alpha \gamma e^{\gamma \phi} \phi'{}^3}{2(r + 4 \alpha \gamma e^{\gamma \phi} \phi')} + {\cal O}\left(\dfrac{1}{X'} \right)\,,
\end{equation}
in which the leading term is $X'$. To obtain detailed forms of $X'(r_h)$, $\phi_h$, and $\phi'_h$, Eq.~(\ref{eq:ode}) can be expanded at the near-horizon limit after inserting Eq.~(\ref{eq:eyh}). Then, the leading terms of Eq.~\eqref{eq:ode} are
\begin{align} 
\phi'' =& \dfrac{(r + 4 \alpha \gamma e^{\gamma \phi} \phi')(r^3 \phi' + 12 \alpha \gamma e^{\gamma \phi} + 4r^2 \alpha \gamma e^{\gamma \phi} \phi'{}^2)}{r^3(r + 4 \alpha\gamma e^{\gamma \phi} \phi') - 96 \alpha^2 \gamma^2 e^{2\gamma \phi}} X' + {\cal O}(1)\,,\label{eq:ode21} \\ 
X'' =& \dfrac{r^4 + 8r^3 \alpha \gamma e^{\gamma \phi} \phi' - 48\alpha^2 \gamma^2 e^{2\gamma \phi} + 16r^2 \alpha^2 \gamma^2 e^{2\gamma \phi} \phi'{}^2 }{r^3(r + 4 \alpha\gamma e^{\gamma \phi} \phi') - 96 \alpha^2 \gamma^2 e^{2\gamma \phi}}X'{}^2 + {\cal O}(X')\,.\label{eq:ode22}
\end{align}
For $\phi''_h$ to be finite, the factor $(r^3 \phi' + 12 \alpha \gamma e^{\gamma \phi} + 4r^2 \alpha \gamma e^{\gamma \phi} \phi'{}^2)$ in Eq.~(\ref{eq:ode21}) must be assumed to be zero, which simplifies Eqs.~(\ref{eq:ode21}) and (\ref{eq:ode22}) to 
\begin{align}\label{eq:ede23a}
\phi'' = {\mathcal O}(1)\,,\quad X''=X'^2+\mathcal{O}(X')\,,
\end{align}
where the coefficient in front of $X'^2$ goes to unity at the near-horizon limit. Now, the differential equations can be solved to obtain the function $X' = x_1/(r-r_h) + {\cal O}(1)$ at the near-horizon limit, fixing the coefficient $x_1$ to unity as the initial condition. Furthermore, $\phi_h$ and $\phi'_h$ are related at the horizon $r_h$ by the condition for $\phi''_h$ being finite, which is
\begin{equation} \label{eq:phiph}
\phi'_h = - \dfrac{r_h e^{-\gamma \phi_h}}{8 \alpha \gamma} \left( 1 \pm \sqrt{1 - \dfrac{192 \alpha^2 \gamma^2 e^{2\gamma \phi}}{r_h^4}} \right)\,,
\end{equation}
where $\phi_h'$ can be determined by setting $r_h$ and $\phi_h$. Then, in the choice of $r_h$ and $\phi_h$, $\phi'$ should be real. Hence, from Eq.~(\ref{eq:phiph}), possible values of $\phi_h$ should satisfy
\begin{equation} \label{eq:phicon}
\phi_h \leq \dfrac{1}{2\gamma} \log \left( \dfrac{r_h^4}{192 \alpha^2 \gamma^2} \right)\,,
\end{equation}
in which all values of $\phi$ can solve Eq.~(\ref{eq:ode}). The solution for the black hole should satisfy specific $X(r)$, $Y(r)$, and $\phi (r)$ boundary conditions. In the asymptotic region, $r\gg 1$, the flatness of the spacetime is ensured by the form of the metric\cite{Kanti:1995vq,Torii:1996yi,Kanti:1997br,Torii:1998gm}
\begin{eqnarray} \label{eq:exae}
e^{X(r)} &=& 1 - \dfrac{2M}{r} + \cdots, \\ \label{eq:eyae}
e^{Y(r)} &=& 1 + \dfrac{2M}{r} + \cdots, \\ \label{eq:phiae}
\phi(r) &=& \dfrac{Q}{r} + \cdots,
\end{eqnarray}
where $M$ and $Q$ denote the ADM mass and dilaton charge of the dilatonic black hole. Note that the asymptotic form of the dilaton field is proportional to $1/r$. This form is different from the logarithmic forms of dilaton fields in other models such as\cite{Cai:1997ii,Hendi:2010gq,Hendi:2015xya}, where there is no Gauss-Bonnet term. The form of the dilaton field will depend on the existence of the Gauss-Bonnet term and choice of the metric ansatz. $\alpha$ and $\gamma$ are fixed to obtain the dilatonic black hole solution. In this case, each value of $r_h$ gives a range of $\phi_h$ satisfying Eq.~\eqref{eq:phicon}. A solution can then be obtained for any initial value set $(r_h,\phi_h)$. However, the dilaton field of the real solution is zero in the asymptotic region, as shown in Eq.~\eqref{eq:phiae}. The real dilaton solution is the only one for given a set ($\alpha$, $\gamma$, $r_h$). With $(\phi_h, r_h)$, $\phi'_h$ can be obtained from Eq.~\eqref{eq:phiph}, where the positive sign is selected to retrieve the asymptotic behavior of the dilaton field. The value of $X'$ at the horizon is given by $X'_h = 1/\epsilon$, where $\epsilon=10^{-8}$ is introduced to avoid the initial singularity from Eq.~(\ref{eq:ede23a}). Since the initial value of $e^{Y(r)}$ can be obtained from Eq.~(\ref{eq:ey}), the initial conditions for the equations of motion are only
\begin{eqnarray} \label{eq:ic}
\phi'_h = - \dfrac{e^{-\gamma \phi_h} r_h}{8 \alpha \gamma} \left( 1 + \sqrt{1 - \dfrac{192 \alpha^2 \gamma^2 e^{2\gamma \phi}}{r_h^4}} \right)\,,\quad X'_h = \dfrac{1}{\epsilon}\,.
\end{eqnarray}
To find the dilatonic black hole solution, we used one of the Runge-Kutta methods with a specific parameter set, the Dormand-Prince method. The equations of motion are solved from $r_h + \epsilon$ to $r_{{max}} = 10^6$, which was considered to be infinity. After the equations are solved, we obtained numerical functions for $X'(r)$, $Y(t)$, and $\phi(r)$. Then, the numerical form of $X(r)$ can be obtained by numerical integration of $X(r)$ with respect to $r$ from $r_h+\epsilon$ to $r_{max}$. The ADM masses $M$ are obtained by fitting Eq.~\eqref{eq:exae} to the solution. The dilatonic black hole solutions are obtained for given values of $\gamma$ as shown in Fig.~\ref{fig:solgamma}, which are the same as the dilatonic black hole solutions reported previously\cite{Kanti:1995vq,Torii:1996yi,Kanti:1997br,Torii:1998gm}.
The mass of the dilatonic black hole $M$ increases as $r_h$ increases. This increase is evident because the mass inside the horizon is proportional to its length in Eq.~(\ref{eq:mass101}). However, for a small horizon, the mass of the black hole is bounded at the minimum mass $M_{min}$ and is two-valued for a given horizon, as shown in Fig.~\ref{fig:solgamma}~(a) and (b), which is an important cause of the interesting behavior of the upper limit of the gravitational radiation. The effect of the dilaton hair becomes important in a black hole with a small mass, which has less gravity than a black hole with a greater mass. This effect originates from the small mass of the black hole having a long hair. Hence, the behavior depends on the coupling $\gamma$ and disappears for values less than $\gamma=1.29859$, as shown in Fig.~\ref{fig:solgamma}~(c). The overall behavior of the metric component is shown in Fig.~\ref{fig:solgamma2}~(a), where the solution can be recognized as that of a Schwarzschild black hole coupled with a dilaton hair. As the mass of the dilatonic black hole increases, the black hole more closely approximates a Schwarzschild black hole, so the effect of the dilaton hair becomes a smaller for more massive dilatonic black holes. In the solution shown in Fig.~\ref{fig:solgamma2}, the large $r_h$ becomes a black hole for a small value of the dilaton hair strength $\phi_h$. Then, in the asymptotic region, $\phi_h$ vanishes, as can be chosen by selecting an appropriate solution to the equations of motion.

The mass of the dilatonic black hole consists of the mass of Schwarzschild black hole $M_{BH}$ and the dilaton hair contribution $M_d$. This characteristic can be seen from the metric component $g^{rr}=e^{-Y(r)}$. When we consider this $g^{rr}$ component to be $e^{Y(r)}=1-\frac{2M(r)}{r}$, where the mass function $M(r)$ is the mass inside a sphere of radius $r$, the mass function should satisfy the boundary conditions,
\begin{align}\label{eq:mass101}
M(r_h)=\frac{r_h}{2}\,,\quad  \lim_{r\rightarrow \infty} M(r)=M\,,
\end{align}
which implies that the ADM mass consists of two contributions, one each from inside and outside the outer horizon. The mass inside the dilatonic black hole is the same as that of a Schwarzschild black hole. Hence, we call it the Schwarzschild mass $M_{BH}=\frac{r_h}{2}$. Since the dilaton field is only one component outside the horizon, the difference between $M$ and $M_{BH}$ is the mass contribution of the dilaton hair stretched outside of the black hole. Therefore, this difference can be set equal to $M_d$. Then, the mass of the dilatonic black hole can be written as\cite{Sudarsky:2002mk},
\begin{equation}\label{eq:mass100}
M = M_{BH}(r_h) + M_{d}.
\end{equation}
Thus, because the mass of the dilatonic black hole is an arithmetic sum of two contributions, it can be treated separately.

In this work, the analysis focuses on the thermal upper bound of the radiation in which the entropy of the black hole plays an important role. The entropy of a dilatonic black hole is given by\cite{Torii:1996yi}
\begin{align}\label{eq:entrop:en1}
S_{BH}=\pi r_h^2-16\alpha \pi e^{\gamma \phi_h}\,,
\end{align}
where the first term is the contribution of the horizon area of the black hole similar to the Bekenstein-Hawking entropy, and the second term is the correction to the Gauss-Bonnet term. The entropy has two limits related to parameters of the black hole solution. In the limit in which $\alpha$ tends to zero, the Gauss-Bonnet term is removed from the action Eq.~(\ref{eq:lagrangian}). According to the no-hair theorem, the metric becomes that of a Schwarzschild black hole in Einstein gravity, and then the area term only remains in Eq.~(\ref{eq:entrop:en1}). The other limit is that in which $\gamma$ tends to zero. In this limit, $\phi_h$ is negligible in Eq.~(\ref{eq:entrop:en1}), and the action becomes that of Einstein gravity coupled with the Gauss-Bonnet term in which there is no dilaton hair. Although the Gauss-Bonnet term still exists, the metric is the same as that of a Schwarzschild black hole, but the entropy is given by
\begin{align}\label{eq:entrop:en2}
S_{BH}=\pi r_h^2-16\alpha\pi\,,
\end{align}
which has a constant contribution from the Gauss-Bonnet term. This feature caused difficulties in this analysis, which will be discussed in the following section. Therefore, we expect that the effect of the dilaton hair nontrivially appears in the gravitational radiation between dilatonic black holes. In addition, the radiation includes the energy from the dilaton hair released in the process.


\section{Upper Limit of Radiation under Collision of Dilatonic Black Holes}\label{sec3}

To find the upper limit of the gravitational radiation released in a dilatonic black hole collision, we define the initial and final states of the process. Then, the limit is obtained using thermodynamic preference between states, and the effect of the dilaton hair in the collisions is determined.

\subsection{Analytical Approach to the Collision}

We consider the initial state to be one with two dilaton black holes separated far from each other in flat spacetime. Hence, the interactions between them are considered to be negligible. In the initial state, one black hole is defined as having mass $M_1$, horizon $r_1$, and dilaton field strength $\phi_1$, while the other had $M_2$, $r_2$, and $\phi_2$. The total mass of the dilaton black hole $M_{tot}$ includes the contribution of the dilaton field determined by $\phi_1$ and $\phi_2$, making the total mass of the initial state
\begin{align}
M_{tot}=M_1(r_1)+M_2(r_2)\,.
\end{align}
In the final state, we consider the two black holes to merge into a dilatonic black hole with gravitational radiation released in the process. The energy $M_{r}$ released as radiation is defined by denoting the final black hole parameters as $M_f$, $r_f$, and $\phi_f$. In this situation, the conservation of the total mass of the final state can be expressed as
\begin{eqnarray}\label{eq:radmass}
M_{tot}=M_f(r_1, r_2)+M_{r}(r_1,r_2)\,,
\end{eqnarray}
where the minimum mass of the final black hole can be obtained from the inequality of the entropies of the initial and final black holes, $A_{initial}$ and $A_{final}$, respectively. The horizon area of the final black hole should be larger than the sum of the areas of the initial black holes according to the second law of thermodynamics\cite{Hawking:1971tu,Wald:1972sz}. Then,
\begin{eqnarray}\label{eq:inequality01}
A_{initial}=4\pi r_1^2 + 4\pi r_2^2\leq 4\pi r_f^2=A_{final}\,,
\end{eqnarray}
where the entropies of radiation and turbulence have been assumed to be negligible, since the entropy of the radiation is less than that of the black holes and is very small in actual observations\cite{Hawking:1971tu}. In actual observations, the radiation is about 5\%\cite{Abbott:2016blz,Abbott:2016nmj,Abbott:2017vtc}, so the contributions to the entropies of radiation and turbulence can be assumed to be sufficiently small compared with those of black holes in the initial state. The minimum value of the horizon $r_{f,min}$ can be obtained from the equality in Eq.~(\ref{eq:inequality01}). At $r_{f,min}$, the mass of the final black hole is also a minimum, $M_{f,min}$. With this minimum mass, the radiation energy is the maximum of $M_{rad}$, which is the upper limit of the radiation from Eq.~(eq:radmass). Therefore, the limit can be expressed as
\begin{align}
M_{rad}=M_{tot}-M_{f,min}=M_1+M_2-M_{f,min}\,,
\end{align} 
where the masses depend on $r_1$ and $r_2$, as shown in Fig.~\ref{fig:solgamma}, so their behaviors are nonlinear.

Since a solution for two interacting dilatonic black holes remains to be obtained in the action in Eq.~(\ref{eq:lagrangian}), it is necessary to assume the form of the entropy correction in two dilatonic black holes to describe the increase in entropy between the initial and final states. We focus on the correction term in Eq.~(\ref{eq:entrop:en1}), which is from the Gauss-Bonnet term of the action in Eq.~(\ref{eq:lagrangian}). Hence, we assume the entropy correction to be the same in the initial and final states, causing the corrections to cancel each other. Therefore, the inequality in Eq.~(\ref{eq:inequality01}) can be reduced to the area theorem of black holes, because the correction term results from the Gauss-Bonnet term in the action in Eq.~(\ref{eq:inequality01}). This assumption also gives consistent results for arbitrary values of $\gamma$, both zero and nonzero.

At $\gamma=0$, the action of Eq.~(\ref{eq:lagrangian}) becomes the Einstein gravity coupled with the Gauss-Bonnet term. As a topological term, the Gauss-Bonnet term does not change the equations of motion, because it is removed as a total derivative term in the equations of motion, so it cannot affect the dynamics. However, the entropy in Eq.~(\ref{eq:entrop:en2}) has a constant correction term, so it gives different results from Einstein gravity for radiation released during collisions. This difference originates from the definition of the initial state. Irrespective of how far the black holes are from each other, the initial black holes are to be just one solution of the action (\ref{eq:lagrangian}), so the correction term can also be considered once. Then, the entropy increase that occurs due to the collision can be expressed as
\begin{eqnarray}
S_{i}=\pi r_1^2+\pi r_2^2-16\alpha \pi \leq \pi r_f^2-16\alpha \pi=S_{f}\,,
\end{eqnarray}
which gives the same result as the Einstein gravity. If we consider the correction twice, it may be the result obtained by summing up two different action at $\gamma=0$.
\begin{align}
{\cal L} = {\cal L}_1+ {\cal L}_2=\dfrac{R_1}{2} + R^2_{1GB}+\dfrac{R_2}{2} + R^2_{2GB}\,,
\end{align}
where the indices $1$ and $2$ indicate the first and second black holes, respectively. Because the Lagrangians ${\cal L}_1$ and ${\cal L}_2$ have Schwarzschild black hole solutions and correction terms, the sum of their entropies is twice that of the value $-16\alpha \pi$. However, this result is the sum of black hole entropies existing in two different spacetimes $1$ and $2$, hence this case can be ruled out. In the solution for a single black hole system, the action would be from Eq.~(\ref{eq:lagrangian})
\begin{align}\label{eq:actionjgb}
{\cal L} = \dfrac{R}{2} + R^2_{GB}\,,
\end{align}
where two black holes far from each other should be a solution to Eq.~(\ref{eq:actionjgb}). Thus, the contribution of the Gauss-Bonnet term may be added once to the total entropy. In addition, this is consistent with Einstein gravity, both with and without the Gauss-Bonnet term.

If this assumption is generalized to $\gamma \ne 0$, the correction term may be essentially the same in the initial and final states for consistency with the $\gamma=0$ case. Since the black hole solution depends on $r_h$, the initial and final black hole entropies can be expressed as an inequality:
\begin{align}\label{eq:en:inq1}
\pi r_1^2 + \pi r_2^2 + c_i(r_1,r_2) \leq \pi r_f^2 -16\alpha \pi e^{\gamma \phi_f}\,.
\end{align}
For the $\gamma=0$ case, as $\lim \gamma -> 0$, $c_i$ and $-16\alpha \pi e^{\gamma \phi_f}$ should converge to $-16\alpha \pi$. Furthermore, in the massive limit, $r_1,\,\,r_2\gg 1$, a dilatonic black hole tends to a Schwarzschild black hole. In this case, $c_i$ and $-16\alpha \pi e^{\gamma \phi_f}$ should also converge to $-16\alpha \pi$. The exact values of the corrections $c_i$ remain unknown, but they might be coincident with each other in these limits. We then checked whether $c_i$ can be approximately equal to $-16\alpha \pi e^{\gamma \phi_f}$. As $r_h$ increases, $\phi_h$ rapidly decreases, as shown in Fig.~\ref{fig:solgamma}, so an asymptotic observer can observe that the dilaton mass $M_d$ is slightly greater in the initial state than in the final state since $M_d$ is proportional to $\phi_h$. However, the areas of the initial black holes can be stretched by each other, so these two opposing contributions can be expected to cancel each other. Thus, it can be assumed that $c_i\approx-16\alpha \pi e^{\gamma \phi_f}$, which is consistent with the $\gamma=0$ or $M \gg 1$ case. This result is expected based on our assumption, and the exact value must be calculated, which will be done in further work. The contributions of the correction terms cancel out in Eq.~(\ref{eq:en:inq1}), yielding Eq.~(\ref{eq:inequality01}). Using Eq.~(\ref{eq:inequality01}), the limiting mass of the final black hole whose horizon is
\begin{eqnarray}\label{mass:min1}
r_{f,min}=\sqrt{r_1^2+r_2^2}\,,
\end{eqnarray}
which can be obtained, and the limiting amount of radiation will be released when the final black hole is synthesized with the minimum mass given by Eq.~(\ref{eq:radmass}). The dilaton field strength of the final black hole satisfies
\begin{equation} \label{eq:phicon2}
\phi_h \leq \dfrac{1}{2\gamma} \log \left( \dfrac{(r_1^2+r_2^2)^2}{192 \alpha^2 \gamma^2} \right)\,.
\end{equation}
where a real black hole solution satisfying the boundary condition in the asymptotic region, $\lim_{r\rightarrow \infty} \phi(r)=0$, should be found. The difference between the masses of the initial and final black holes is the released radiation energy. The maximum radiation released in the given conditions is the upper limit that is thermodynamically allowed. The detailed behavior of the radiation will depend on parameters such as the horizon and dilaton field strength, as illustrated in the following sections.

\subsection{Upper Limit of Radiation with Dilaton Field}

\begin{figure}[h]
\centering\subfigure[{\small Overall behaviors of the limits of the radiation in $\alpha=1/16$ and $\gamma=\sqrt{2}$.}]
{\includegraphics[scale=0.80,keepaspectratio]{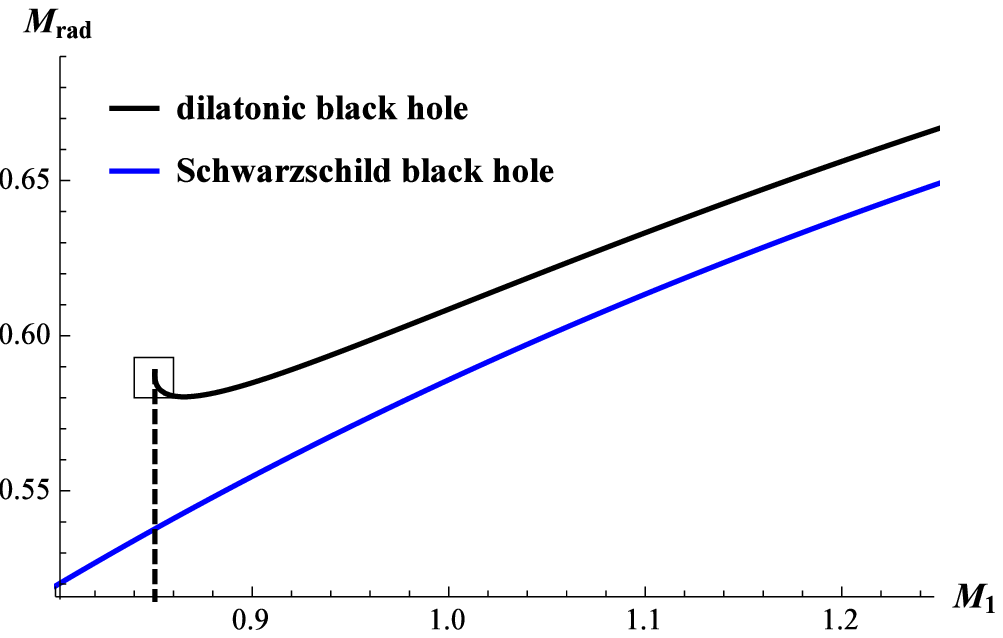}}
\quad
\centering\subfigure[{\small The limit of the radiation at the small mass in $\alpha=1/16$ and $\gamma=\sqrt{2}$.}]
{\includegraphics[scale=0.80,keepaspectratio]{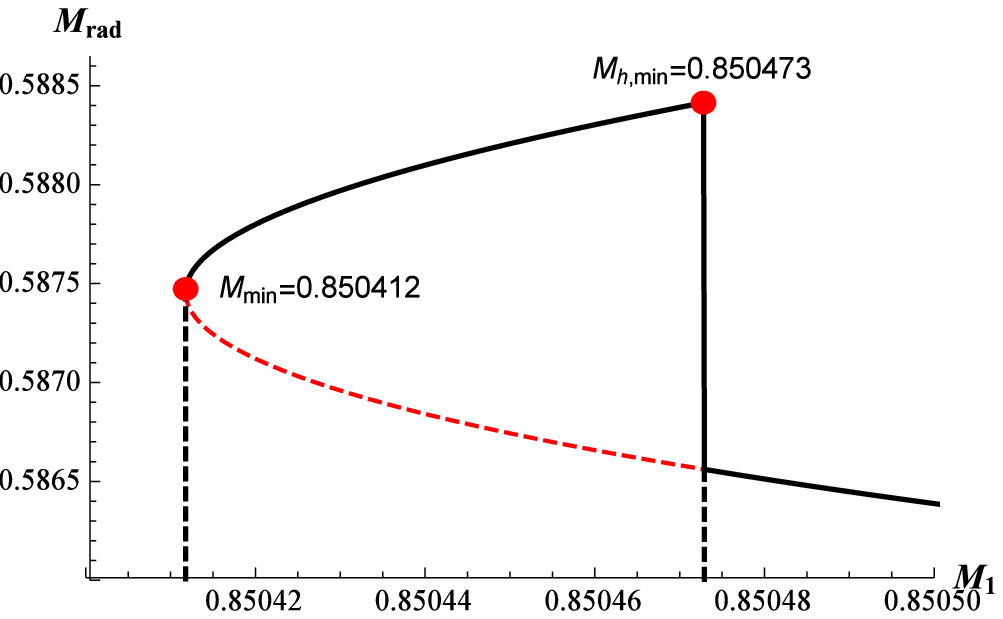}}
\caption{{\small The upper limits of the radiation for the collision of two dilatonic black holes.}}
\label{fig:mrad1}
\end{figure}

\begin{figure}[h]
\centering\subfigure[{\small The limit of the radiation in $\alpha=1/16$ and $\gamma=1.35$.}]
{\includegraphics[scale=0.80,keepaspectratio]{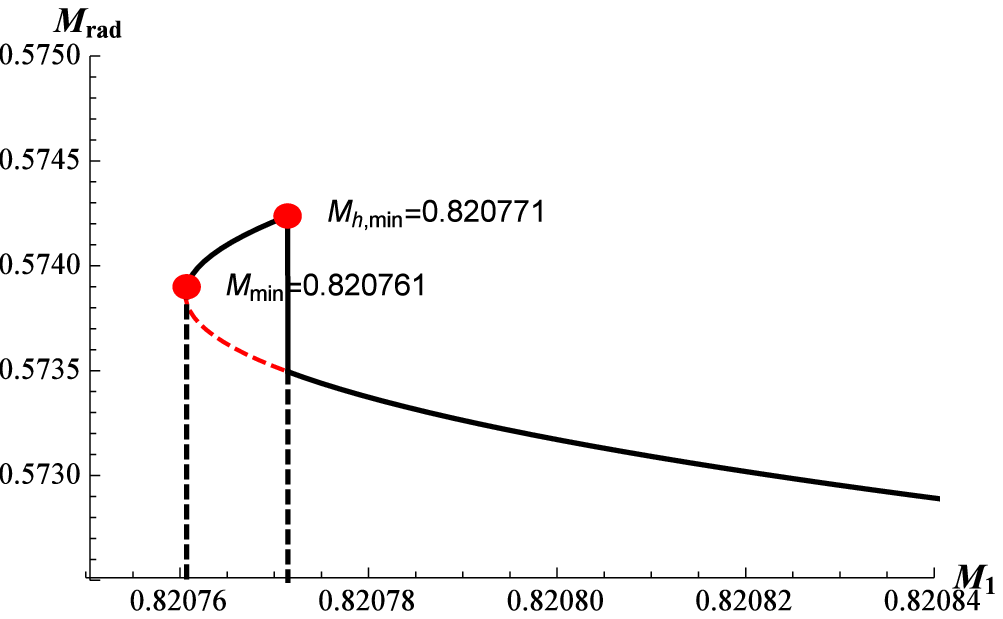}}
\quad
\centering\subfigure[{\small The limit of the radiation in $\alpha=1/16$ and $\gamma=1.29859$.}]
{\includegraphics[scale=0.80,keepaspectratio]{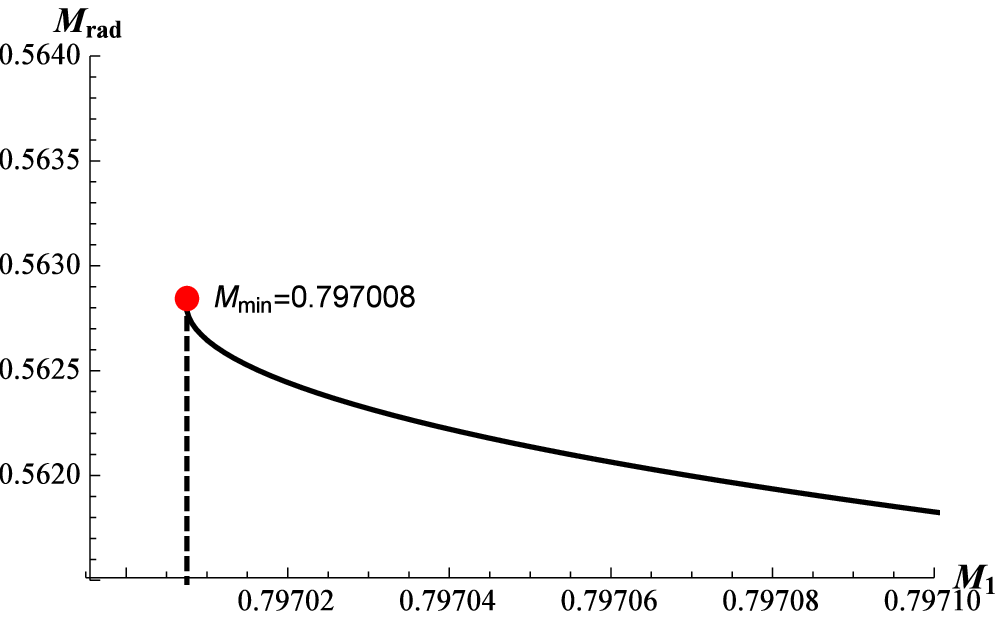}}
\caption{{\small The upper limits of the collision of two dilatonic black hole for different values of $\gamma$.}}
\label{fig:gamma2}
\end{figure}

\begin{figure}[h]
\centering\subfigure[{\small The upper limit of the mass of the dilaton hair radiated out in $\alpha=1/16$ and $\gamma=\sqrt{2}$.}]
{\includegraphics[scale=0.80,keepaspectratio]{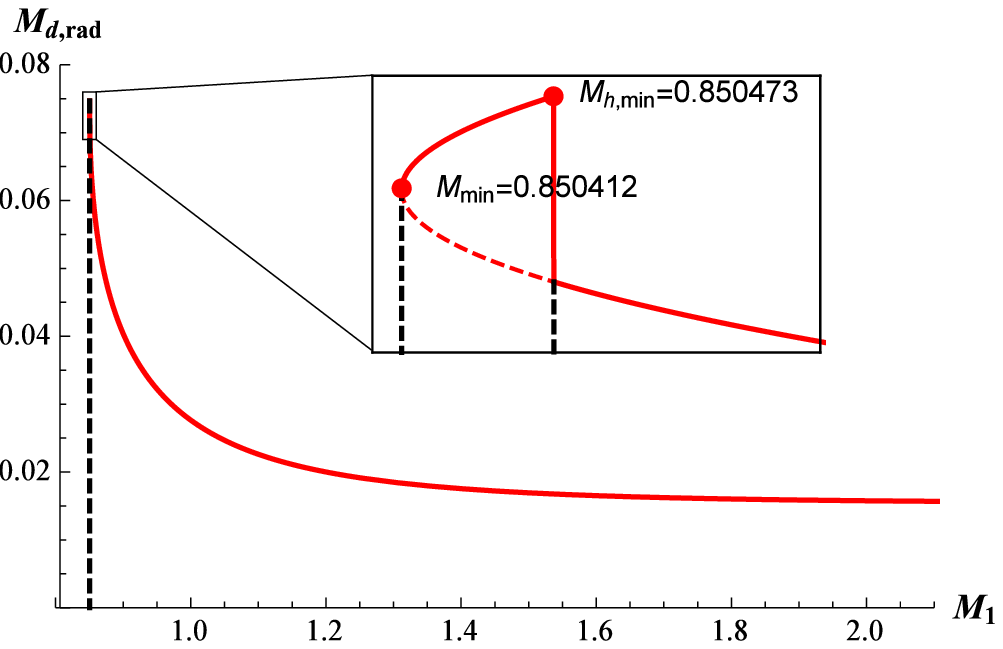}}
\quad
\centering\subfigure[{\small The ratio of between the initial mass and upper limit in $\alpha=1/16$ and $\gamma=\sqrt{2}$.}]
{\includegraphics[scale=0.80,keepaspectratio]{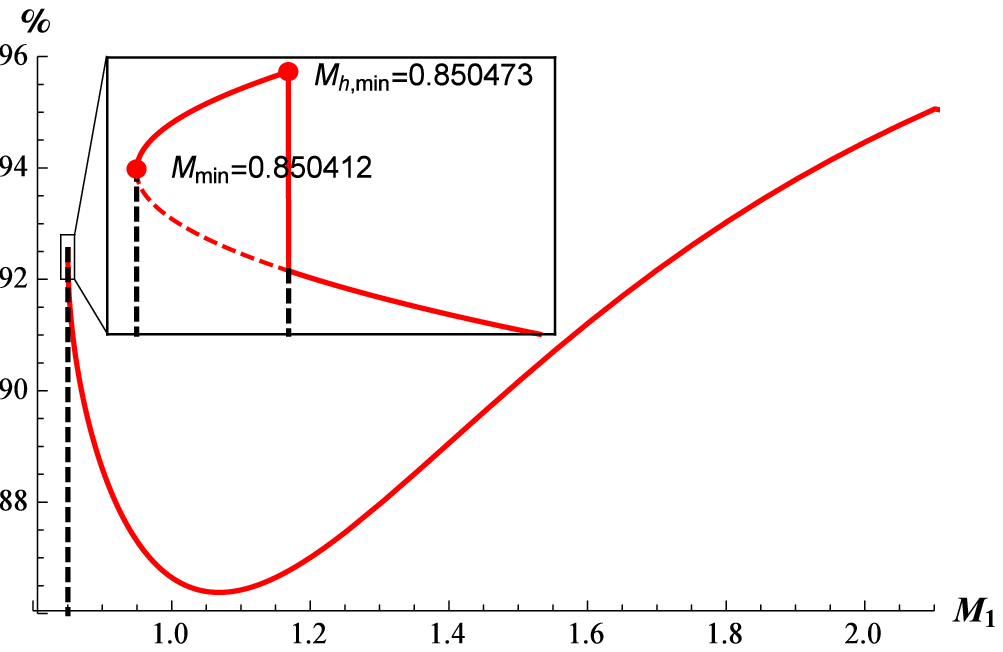}}
\caption{{\small The upper limits of the radiation of a dilaton hair for the collision of two dilatonic black holes.}}
\label{fig:f5dilatonrad}
\end{figure}

The upper limit of the gravitational radiation released by dilatonic black holes with the Gauss-Bonnet term is dependent on the masses and dilaton field strengths of the black holes. For a black hole with a large mass, the minimum horizon of the final black hole is given by Eq.~(\ref{mass:min1}) and is proportional to its minimum mass. As its mass increases, the limit of the radiation also increases, as shown in Fig.~\ref{fig:mrad1}\,(a), which depicts the relation between the mass of the first black hole $M_1$ and the limit of the radiation energy when the parameters of the second black hole are fixed as follows: $r_2 = 2$, $M_2 = 1.013263$, and $\phi_2 = 0.19192$. Note that these values are the same from Figs.~\ref{fig:mrad1} to \ref{fig:ratiop1}.
The radiation energy includes the contribution from the dilaton hair, so the limit of the radiation from a dilatonic black hole is greater than that from a Schwarzschild black hole. However, in a dilatonic black hole with a small mass, the limit of the radiation begins at $M_{min}$ for the first black hole, which has the minimum value in $M_{rad}$. This limiting value results from the solution for a dilatonic black hole with the minimum mass $M_{min}$, which is shown in Fig.~\ref{fig:solgamma}. Hence, diatonic black holes exhibit behaviors very different from those of Schwarzschild black holes. In Fig.~\ref{fig:mrad1} \,(a), the limit for a dilatonic black hole begins at $M_{min}$ for the first black hole, because the black hole solution does not exist for small values of $r_h$ in Fig.~\ref{fig:solgamma}. In addition, the limit of radiation has a minimum value and a discontinuity at $M_{h,min}$, as shown in Fig.~\ref{fig:mrad1} \,(b), since the black hole solution has a minimum value and two solutions for a given mass $M$, as shown in Fig.~\ref{fig:solgamma}. Since the mass of the final black hole has a value in the range of $\sqrt{r_1^2+r_2^2}$, which is sufficiently large compared to the range containing the two solution values, the limit of the radiation depends on the behaviors of $M_1(r_1)$ rather than $M_{f,min}$. In the range of the two solutions, the mass of the first black hole $M_1$ is large at $r_{1,min}$, the smallest radius, so $M_{rad}$ becomes large, as shown in Fig.~\ref{fig:mrad1}\,(b). For a given mass $M_1$,
\begin{eqnarray}
M_{min}\leq M_1(r_{1,min})\,,
\end{eqnarray}  
where the equality is satisfied for small $\gamma$. Hence, the radiation limit has a discontinuity at $M_{h,min}$ in the range of the two solutions.
The thermally allowed upper limit of the radiation is represented by the black line in Fig.~\ref{fig:mrad1}\,(b). $M_{h,min}$ depends on $\gamma$ in the range of the two solutions, as shown in Fig.~\ref{fig:gamma2}. For large $\gamma$, the discontinuity persists, as shown in Fig.~\ref{fig:gamma2}\,(a), but it approximates the minimum value of $M_1$ for small $\gamma$. Finally, for $\gamma = 1.29859$, the points overlap with each other, and there is no discontinuous upper limit, as shown in Fig.~\ref{fig:gamma2}\,(b), which has been observed for dilatonic black holes but not Schwarzschild black holes.

The mass of a dilatonic black hole, such as $M_1$, $M_2$, or $M_{f}$, includes the dilaton mass. Hence, $M_{rad}$ also includes a contribution from the dilaton hair. Therefore, the mass of the dilaton hair that can be released due to the collision can be determined. The mass of the black hole can be expressed in terms of its own mass $M_{BH}$ and the dilaton mass $M_{d}$. $M_{BH}$ is half of $r_h$ from Eq.~(\ref{eq:entrop:en1}):
\begin{eqnarray}
M=M_{BH}+M_d=\frac{r_h}{2}+M_d\,.
\end{eqnarray}
Therefore, the energy of the dilaton hair released as radiation $M_{d,rad}$ can be obtained as shown in Fig.~\ref{fig:f5dilatonrad}.
The overall behavior of the upper limit of the dilaton hair radiation is presented in Fig.~\ref{fig:f5dilatonrad}~(a). Since the mass of the hair in the dilatonic black hole is very small, the dilaton hair released due to radiation is also very small compared to the total mass of the initial state. As the mass of the black hole increases, the dilatonic black hole approximates a Schwarzschild black hole. Therefore, the dilaton contribution to the radiation is large when the mass of the initial state is small. The dilaton contribution to the radiation is the largest at $M_{h,min}$ and starts at the minimum mass $M_{min}$, as shown in Fig~\ref{fig:mrad1}. The point of discontinuity disappears for small values of $\gamma$. This is also identical to the limit of the radiation. In addition, throughout the process, most of the dilaton hair is released, as shown in Fig.~\ref{fig:f5dilatonrad}~(b), which shows the ratio of the dilaton contribution to the radiation to the dilaton mass of the initial black hole,
\begin{eqnarray}
M_{d,rad}(\%)=\frac{(M_1+M_2-\frac{r_1}{2}-\frac{r_2}{2})-(M_f-\frac{r_f}{2})}{M_1+M_2-\frac{r_1}{2}-\frac{r_2}{2}}\,.
\end{eqnarray} 
This ratio is approximately 90\%. Hence, most of the dilaton hair in the initial state is radiated out during the collision process. Due to the dilaton effect, the radiation limit still has a point of discontinuity $M_{h,min}$, but $M_{d,rad}(\%)$, which is close to $100\%$ in a massive black hole, has no maximum. As the mass increases, a dilatonic black hole more closely approximates a Schwarzschild black hole, so the abovementioned ratio becomes small for a massive final black hole. If the dilatonic black hole is massive, the dilaton hair is also released in larger quantities, which results in the increases shown in Fig.~\ref{fig:f5dilatonrad}~(b). Therefore, since the mass of the dilatonic black hole increases due to the collision, the final black hole is similar to a Schwarzschild black hole, and no dilaton hair can be detected.

\subsection{Notes on GW150914 and GW151226}

\begin{figure}[h]
\begin{center}
\includegraphics[scale=0.80,keepaspectratio]{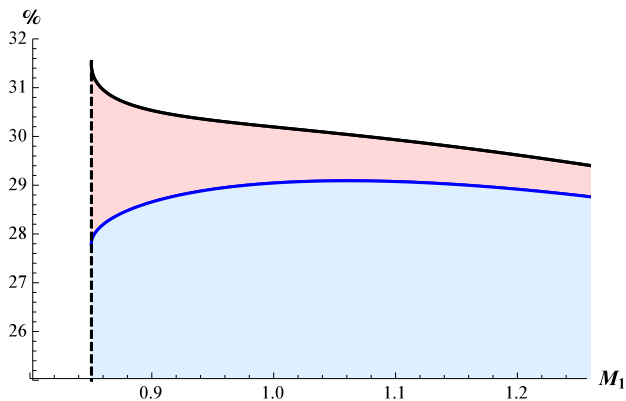}
\end{center}
\caption{{\small The upper limit of the radiation under the collision in $\alpha=1/16$ and $\gamma=\sqrt{2}$. The total energy of the radiation is a black line, and the radiated energy from the mass of inside of the horizon is given as a blue line.}}
\label{fig:ratiop1}
\end{figure}

The radiation released with respect to the total mass of the initial state can be obtained and divided into two parts as
\begin{align}
M_{rad}(\%)=\frac{(M_1+M_2)-M_f}{M_1+M_2}=\frac{(\frac{r_1}{2}+\frac{r_2}{2})-\frac{r_f}{2}}{M_1+M_2}+\frac{(M_1-\frac{r_1}{2}+M_2-\frac{r_2}{2})-(M_f-\frac{r_f}{2})}{M_1+M_2}\,,
\end{align}
where the first term is the contribution of the black hole mass, and the second term is that of the dilaton hair. The mass released due to gravitational radiation is thermally limited at approximately 30\% of the total mass of the initial state, as shown in Fig.~\ref{fig:ratiop1}. Most of the radiation is from the black hole mass shown in blue.
However, a dilaton black hole has a contribution from the dilaton hair, which is given by the contribution of the dilaton hair less than 10\% of the upper limit of the total radiation shown in red in Fig.~\ref{fig:ratiop1}. The ratio of the contribution of the dilaton hair to the total mass of the initial black holes is larger when the mass is smaller, which can be seen from the solution of the black hole.

If these results are applied to GW150914 and GW121226, which were detected by LIGO\cite{Abbott:2016blz,Abbott:2016nmj}, the upper limit is consistent with the experimental observation. GW150914 was generated by the merger of two black holes with masses of $39M_\odot$ and $32M_\odot$ in a detector frame considered with a redshift $z=0.09$\cite{Krolak:1987ofj,Ade:2015xua}. The final state is a black hole with a mass of $68 M_\odot$ and a $3M_\odot$ gravitational wave. In this case, the radiation is approximately $4\%$ of the total mass of the initial state. In a similar way, for GW151226, the radiation is also approximately $4\%$ of the total mass. Therefore, the ratios obtained for the gravitational waves detected by LIGO are near the upper limit of the radiation obtained from the thermodynamics calculations. If the ratio of the dilaton hair is assumed to be the same in both the upper limit and detection, the contribution of the dilation hair can comprise up to approximately $10\%$ of the radiation, so that the real contribution is $0.4\%$ of the total amount of radiation, which is significant considering that the radiation energies are $0.3 M_\odot$ for GW150914 and $0.1 M_\odot$ for GW151226, since the masses are in units of solar mass. However, this is based on the maximum ratio, so the exact contribution may be much less than $10\%$.

\section{Summary}\label{sec5}
We investigated the upper limit of the gravitational radiation released in a dilaton black hole collision using the Gauss-Bonnet term. The solutions for dilatonic black holes were obtained numerically. In the solution, the total mass consists of the black hole and dilaton hair masses. As the black hole horizon becomes larger, the total mass increases, but there are two black hole solutions for a given radius $r_h$ when smaller masses are involved. This feature plays an important role in radiation emission. To determine the upper limit of the radiation that is thermally allowed, we assumed the dilatonic black holes to be far apart from one another and a head-on collision between them to produce the final black hole. The mass difference between the initial and final black holes was determined based on the energy of the gravitational radiation. Since such collisions are irreversible, the entropy of the final black hole should be larger than that of the initial state. In addition, we assumed the correction terms in the entropies to cancel each other in the initial and final states, because such collisions occur in one spacetime and each correction term can be expected to contribute the same value on both sides. Using this thermal preference, the upper limit of the radiation energy in the collision can be obtained.

The upper limit is larger than that of a Schwarzschild black hole, since the radiation includes not only the mass of the black hole, but also its dilaton hair. The upper limit starts at the minimum black hole mass and is proportional to the black hole mass; however, when the mass is small, a point of discontinuity originates from the two solutions for a given mass. The point of discontinuity depends on $\gamma$ and vanishes for $\gamma<1.29859$. Due to the collision, the dilaton hair can be radiated out. Most of the mass from the dilaton hair in the initial state, approximately $90\%$ of the initial hair mass, is released, so the final black hole has a very small amount of hair compared with the initial one. In the total mass of the initial state, the upper limit of the radiation is approximately $30\%$, and the radiation of the dilaton hair contributes a maximum of $10\%$ to the initial mass. Therefore, the hair contribution should be considered in gravitational radiation. We also discussed the possible mass of the hair radiated out, such as in GW150914 and GW151226 detected by LIGO.

\vspace{10pt} 

{\bf Acknowledgments}

This work was supported by the faculty research fund of Sejong University in 2016. BG was supported by Basic Science Research Program through the National Research Foundation of Korea(NRF) funded by the Ministry of Science, ICT \& Future Planning(NRF-2015R1C1A1A02037523). DR was supported by the Korea Ministry of Education, Science and Technology, Gyeongsangbuk-Do and Pohang City.

\appendix
\section{Appendix} \label{sec:A}

\begin{eqnarray} \nonumber
N_1 &=& -2 r X'{}^2 \left(e^{Y} r-8 e^{\gamma  \phi} \alpha  \gamma  \phi'\right) \left(e^{Y} r+4 e^{\gamma  \phi} \left(-3+e^{Y}\right) \alpha  \gamma  \phi'\right)^2 
\\ \nonumber
&& 
+2 e^{Y} \big(-32 e^{2 \gamma  \phi} \left(12-7 e^{Y}+e^{2 Y}\right) r^2 \alpha^2 \gamma^2 \phi'{}^4 
\\ \nonumber
&& 
+8 e^{\gamma  \phi} r \alpha  \gamma  \left(7 e^{Y} r^2-e^{2 Y} r^2+24 e^{\gamma  \phi} \alpha  \gamma^2-48 e^{Y+\gamma  \phi} \alpha  \gamma^2+24 e^{2 Y+\gamma  \phi} \alpha  \gamma^2\right) \phi'{}^3 
\\ \nonumber
&& 
-\big(2 e^{2 Y} r^4+e^{3 Y} r^4+16 e^{Y+\gamma  \phi} \alpha  \gamma^2 r^2-32 e^{2 Y+\gamma  \phi} \alpha  \gamma^2 r^2+16 e^{3 Y+\gamma  \phi} \alpha  \gamma^2 r^2+192 e^{2 \gamma  \phi} \alpha^2 \gamma^2  
\\ \nonumber
&& 
+ 64 e^{2 (Y+\gamma  \phi)} \alpha^2 \gamma^2 - 256 e^{Y+2 \gamma  \phi} \alpha^2 \gamma^2 \big) \phi'{}^2 -16 e^{Y+\gamma  \phi} \left(-1+e^{2 Y}\right) r \alpha  \gamma  \phi' + 2 e^{3 Y} \left(-1+e^{Y}\right) r^2 \big) 
\\ \nonumber
&& 
+ X'\bigg(-96 e^{2 \gamma  \phi} \left(-1+e^{Y}\right) r \alpha^2 \gamma^2 \left(e^{Y} r^2-16 e^{\gamma  \phi} \alpha  \gamma^2+16 e^{Y+\gamma  \phi} \alpha  \gamma^2\right) \phi'{}^4 
\\ \nonumber
&& 
+ 4 e^{\gamma  \phi} \left(-3 + e^{Y}\right) \alpha  \gamma  \left(e^{2 Y} r^4-32 e^{Y+\gamma  \phi} \alpha  \gamma^2 r^2+32 e^{2 Y+\gamma  \phi} \alpha  \gamma^2 r^2 -384 e^{2 \gamma  \phi} \alpha^2 \gamma^2+128 e^{Y+2 \gamma  \phi} \alpha^2 \gamma^2 \right) \phi'{}^3
\\ \nonumber
&& 
+ e^{Y} r \bigg(e^{2 Y} r^4-32 e^{Y+\gamma  \phi} \alpha  \gamma^2 r^2+32 e^{2 Y+\gamma  \phi} \alpha  \gamma^2 r^2-1344 e^{2 \gamma  \phi} \alpha^2 \gamma^2+64 e^{2 (Y+\gamma \phi)} \alpha^2 \gamma^2
\\
&& 
+512 e^{Y+2 \gamma  \phi} \alpha^2 \gamma^2\bigg) \phi'{}^2 -8 e^{2 Y+\gamma \phi} \left(-15+2 e^{Y}+e^{2 Y}\right) r^2 \alpha \gamma \phi'-2 e^{3 Y} \left(1+e^{Y}\right) r^3 \bigg),
\end{eqnarray}

\begin{eqnarray} \nonumber
N_2 &=& -8 e^{\gamma \phi} r \alpha \gamma \left(-e^{2 Y} \left(-3+e^{Y}\right) r^2-4 e^{Y+\gamma  \phi} \left(9-2 e^{Y}+e^{2 Y}\right) \alpha  \gamma  \phi' r +32 e^{2 \gamma  \phi} \left(3+e^{2 Y}\right) \alpha^2 \gamma^2 \phi'{}^2\right) X'{}^3
\\ \nonumber
&& 
+2 e^{Y} \bigg(16 e^{2 \gamma  \phi} \left(-9+5 e^{Y}\right) r^3 \alpha^2 \gamma^2 \phi'{}^3+2 e^{\gamma  \phi} \alpha  \gamma  \big(13 e^{Y} r^4-3 e^{2 Y} r^4-768 e^{2 \gamma  \phi} \alpha^2 \gamma^2
\\ \nonumber
&& 
+256 e^{Y+2 \gamma  \phi} \alpha^2 \gamma^2 \big) \phi'{}^2 -r \left(e^{2 Y} r^4-96 e^{2 \gamma  \phi} \alpha^2 \gamma^2+32 e^{2 (Y+\gamma  \phi)} \alpha^2 \gamma^2-192 e^{Y+2 \gamma  \phi} \alpha^2 \gamma^2\right) \phi' 
\\ \nonumber
&& 
-4 e^{Y+\gamma  \phi} \left(3+e^{2 Y}\right) r^2 \alpha  \gamma \bigg) X'{}^2 +e^{Y} \bigg(-8 e^{Y+\gamma  \phi} r^3 \alpha  \gamma  \left(3 r^2+32 e^{\gamma  \phi} \alpha  \gamma^2\right) \phi'{}^4
\\ \nonumber
&& 
+r^2 \bigg(e^{2 Y} r^4+16 e^{Y+\gamma  \phi} \alpha  \gamma^2 r^2 +16 e^{2 Y+\gamma  \phi} \alpha  \gamma^2 r^2+192 e^{2 \gamma  \phi} \alpha^2 \gamma^2 +64 e^{2 (Y+\gamma  \phi)} \alpha^2 \gamma^2
\\ \nonumber
&& 
+128 e^{Y+2 \gamma  \phi} \alpha^2 \gamma^2 \bigg) \phi'{}^3 +16 e^{Y+\gamma  \phi} \left(-2+e^{Y}\right) r^3 \alpha  \gamma  \phi'{}^2 +32 e^{Y+\gamma  \phi} \left(-1+e^{Y}\right)^2 r \alpha  \gamma 
\\ \nonumber
&& 
-2 \left(e^{2 Y} r^4+e^{3 Y} r^4+192 e^{2 \gamma  \phi} \alpha^2 \gamma^2+192 e^{2 (Y+\gamma  \phi)} \alpha^2 \gamma^2-384 e^{Y+2 \gamma  \phi} \alpha^2 \gamma^2\right) \phi' \bigg) X'
\\ \nonumber
&& 
-2 e^{2 Y} \big(8 e^{\gamma  \phi} \left(-3+e^{Y}\right) r^4 \alpha  \gamma  \phi'{}^4+r^3 \left(e^{Y} r^2+16 e^{\gamma  \phi} \alpha  \gamma^2-16 e^{Y+\gamma  \phi} \alpha  \gamma^2\right) \phi'{}^3
\\
&& 
+8 e^{\gamma  \phi} \left(-1-e^{Y}+2 e^{2 Y}\right) r^2 \alpha  \gamma  \phi'{}^2+2 e^{Y} \left(-1+e^{Y}\right) r^3 \phi'-16 e^{\gamma  \phi} \left(-1+e^{Y}\right)^2 \alpha  \gamma \big),
\end{eqnarray}

\begin{eqnarray} \nonumber
D &=& 4 r \bigg(8 e^{\gamma  \phi} \left(-1+e^{Y}\right) \alpha  \gamma X' \left(-e^{2 Y} r^2-4 e^{Y+\gamma  \phi} \left(-3+e^{Y}\right) \alpha \gamma \phi' r+48 e^{2 \gamma  \phi} \left(-1+e^{Y}\right) \alpha^2 \gamma^2 \phi'{}^2\right)
\\ \nonumber
&& 
+e^{Y} \big( 4 e^{Y+\gamma  \phi} \left(-5+e^{Y}\right) \alpha \gamma \phi'{}^2 r^3-32 e^{2 \gamma  \phi} \left(-3+e^{Y}\right) \alpha^2 \gamma^2 \phi'{}^3 r^2+8 e^{Y+\gamma \phi} \left(-1+e^{Y}\right)^2 \alpha \gamma r
\\
&& 
+\left(e^{2 Y} r^4-96 e^{2 \gamma \phi} \alpha^2 \gamma^2-96 e^{2 (Y+\gamma  \phi)} \alpha^2 \gamma^2+192 e^{Y+2 \gamma \phi} \alpha^2 \gamma^2 \right) \phi' \big) \bigg).
\end{eqnarray}


\begin{thebibliography}{99}

\bibitem{Abbott:2016blz} 
  B.~P.~Abbott {\it et al.} [LIGO Scientific and Virgo Collaborations],
  Phys.\ Rev.\ Lett.\  {\bf 116}, no. 6, 061102 (2016).

\bibitem{Abbott:2016nmj} 
  B.~P.~Abbott {\it et al.} [LIGO Scientific and Virgo Collaborations],
Phys.\ Rev.\ Lett.\  {\bf 116}, no. 24, 241103 (2016).

\bibitem{Abbott:2017vtc} 
  B.~P.~Abbott {\it et al.} [LIGO Scientific and VIRGO Collaborations],
  Phys.\ Rev.\ Lett.\  {\bf 118}, no. 22, 221101 (2017).

\bibitem{Ruffini:1971bza} 
  R.~Ruffini and J.~A.~Wheeler,
Phys.\ Today {\bf 24}, no. 1, 30 (1971).

\bibitem{Bekenstein:1995un} 
  J.~D.~Bekenstein,
Phys.\ Rev.\ D {\bf 51}, no. 12, R6608 (1995).

\bibitem{Mayo:1996mv} 
  A.~E.~Mayo and J.~D.~Bekenstein,
Phys.\ Rev.\ D {\bf 54}, 5059 (1996).

\bibitem{Gibbons:1982ih} 
  G.~W.~Gibbons,
Nucl.\ Phys.\ B {\bf 207}, 337 (1982).

\bibitem{Gibbons:1987ps} 
  G.~W.~Gibbons and K.~i.~Maeda,
Nucl.\ Phys.\ B {\bf 298}, 741 (1988).

\bibitem{Garfinkle:1990qj} 
  D.~Garfinkle, G.~T.~Horowitz and A.~Strominger,
Phys.\ Rev.\ D {\bf 43}, 3140 (1991)
Erratum: [Phys.\ Rev.\ D {\bf 45}, 3888 (1992)].

\bibitem{Droz:1991cx} 
  S.~Droz, M.~Heusler and N.~Straumann,
Phys.\ Lett.\ B {\bf 268}, 371 (1991).

\bibitem{Lee:1991vy} 
  K.~M.~Lee, V.~P.~Nair and E.~J.~Weinberg,
Phys.\ Rev.\ D {\bf 45}, 2751 (1992).

\bibitem{Breitenlohner:1991aa} 
  P.~Breitenlohner, P.~Forgacs and D.~Maison,
Nucl.\ Phys.\ B {\bf 383}, 357 (1992).

\bibitem{Lavrelashvili:1992cp} 
  G.~V.~Lavrelashvili and D.~Maison,
Phys.\ Lett.\ B {\bf 295}, 67 (1992).

\bibitem{Torii:1993vm} 
  T.~Torii and K.~i.~Maeda,
Phys.\ Rev.\ D {\bf 48}, 1643 (1993).

\bibitem{Breitenlohner:1994di} 
  P.~Breitenlohner, P.~Forgacs and D.~Maison,
Nucl.\ Phys.\ B {\bf 442}, 126 (1995).

\bibitem{Kleihaus:2011tg} 
  B.~Kleihaus, J.~Kunz and E.~Radu,
Phys.\ Rev.\ Lett.\  {\bf 106}, 151104 (2011).

\bibitem{Kleihaus:2014lba} 
  B.~Kleihaus, J.~Kunz and S.~Mojica,
Phys.\ Rev.\ D {\bf 90}, no. 6, 061501 (2014).

\bibitem{Kleihaus:2016rgf} 
  B.~Kleihaus, J.~Kunz and F.~Navarro-Lerida,
Class.\ Quant.\ Grav.\  {\bf 33}, no. 23, 234002 (2016).

\bibitem{Boulware:1986dr} 
  D.~G.~Boulware and S.~Deser,
Phys.\ Lett.\ B {\bf 175}, 409 (1986).

\bibitem{Callan:1988hs} 
  C.~G.~Callan, Jr., R.~C.~Myers and M.~J.~Perry,
Nucl.\ Phys.\ B {\bf 311}, 673 (1989).

\bibitem{Mignemi:1992pm} 
  S.~Mignemi and N.~R.~Stewart,
Phys.\ Lett.\ B {\bf 298}, 299 (1993).

\bibitem{Campbell:1991kz} 
  B.~A.~Campbell, N.~Kaloper and K.~A.~Olive,
Phys.\ Lett.\ B {\bf 285}, 199 (1992).

\bibitem{Campbell:1992hc} 
  B.~A.~Campbell, N.~Kaloper, R.~Madden and K.~A.~Olive,
Nucl.\ Phys.\ B {\bf 399}, 137 (1993).

\bibitem{Callan:1985ia} 
  C.~G.~Callan, Jr., E.~J.~Martinec, M.~J.~Perry and D.~Friedan,
Nucl.\ Phys.\ B {\bf 262}, 593 (1985).

\bibitem{Zwiebach:1985uq} 
  B.~Zwiebach,
Phys.\ Lett.\  {\bf 156B}, 315 (1985).

\bibitem{Gross:1986mw} 
  D.~J.~Gross and J.~H.~Sloan,
Nucl.\ Phys.\ B {\bf 291}, 41 (1987).

\bibitem{Kanti:1995vq}
  P.~Kanti, N.~E.~Mavromatos, J.~Rizos, K.~Tamvakis and E.~Winstanley,
  Phys.\ Rev.\ D {\bf 54}, 5049 (1996).

\bibitem{Torii:1996yi}
  T.~Torii, H.~Yajima and K.~-i.~Maeda,
  Phys.\ Rev.\ D {\bf 55}, 739 (1997).

\bibitem{Kanti:1997br}
  P.~Kanti, N.~E.~Mavromatos, J.~Rizos, K.~Tamvakis and E.~Winstanley,
  Phys.\ Rev.\ D {\bf 57}, 6255 (1998).

\bibitem{Torii:1998gm}
  T.~Torii and K.~-i.~Maeda,
  Phys.\ Rev.\ D {\bf 58}, 084004 (1998).

\bibitem{Kokkotas:2015uma}
 K.~D.~Kokkotas, R.~A.~Konoplya and A.~Zhidenko,
 Phys.\ Rev.\ D {\bf 92}, no. 6, 064022 (2015).

\bibitem{Coleman:1991ku}
  S.~R.~Coleman, J.~Preskill and F.~Wilczek,
  Nucl.\ Phys.\ B {\bf 378}, 175 (1992).

\bibitem{Coleman:1991jf}
  S.~R.~Coleman, J.~Preskill and F.~Wilczek,
  Phys.\ Rev.\ Lett.\  {\bf 67}, 1975 (1991).

\bibitem{Horne:1992zy} 
  J.~H.~Horne and G.~T.~Horowitz,
  Phys.\ Rev.\ D {\bf 46}, 1340 (1992).

\bibitem{Mann:1992yv} 
  R.~B.~Mann,
  Phys.\ Rev.\ D {\bf 47}, 4438 (1993).

\bibitem{Lavrelashvili:1992ia} 
  G.~V.~Lavrelashvili and D.~Maison,
  Nucl.\ Phys.\ B {\bf 410}, 407 (1993).

\bibitem{Gibbons:1994vm} 
  G.~W.~Gibbons, G.~T.~Horowitz and P.~K.~Townsend,
  Class.\ Quant.\ Grav.\  {\bf 12}, 297 (1995).

\bibitem{Cai:1997ii} 
  R.~G.~Cai, J.~Y.~Ji and K.~S.~Soh,
  Phys.\ Rev.\ D {\bf 57}, 6547 (1998).

\bibitem{Cai:2001dz} 
  R.~G.~Cai,
  Phys.\ Rev.\ D {\bf 65}, 084014 (2002).

\bibitem{Cai:2003gr} 
  R.~G.~Cai and Q.~Guo,
  Phys.\ Rev.\ D {\bf 69}, 104025 (2004).

\bibitem{Kim:2007iw} 
  H.~C.~Kim and R.~G.~Cai,
  Phys.\ Rev.\ D {\bf 77}, 024045 (2008).

\bibitem{Goldstein:2009cv} 
  K.~Goldstein, S.~Kachru, S.~Prakash and S.~P.~Trivedi,
  JHEP {\bf 1008}, 078 (2010).

\bibitem{Cai:2013qga} 
  R.~G.~Cai, L.~M.~Cao, L.~Li and R.~Q.~Yang,
  JHEP {\bf 1309}, 005 (2013).

\bibitem{Moura:2006pz} 
  F.~Moura and R.~Schiappa,
  Class.\ Quant.\ Grav.\  {\bf 24}, 361 (2007).

\bibitem{Moura:2012fq}
  F.~Moura,
  Phys.\ Rev.\ D {\bf 87}, no. 4, 044036 (2013).

\bibitem{Penrose:1969pc} 
  R.~Penrose,
  Riv.\ Nuovo Cim.\  {\bf 1}, 252 (1969)
  [Gen.\ Rel.\ Grav.\  {\bf 34}, 1141 (2002)].
  
\bibitem{Wald1974548}
  R.~Wald, 
  Ann. Phys. {\bf 82}, no. 2, 548 (1974).

\bibitem{Jacobson:2009kt} 
  T.~Jacobson and T.~P.~Sotiriou,
  Phys.\ Rev.\ Lett.\  {\bf 103}, 141101 (2009).

\bibitem{Saa:2011wq} 
  A.~Saa and R.~Santarelli,
  Phys.\ Rev.\ D {\bf 84}, 027501 (2011);

\bibitem{Gao:2012ca} 
  S.~Gao and Y.~Zhang,
  Phys.\ Rev.\ D {\bf 87}, no. 4, 044028 (2013).

\bibitem{Barausse:2010ka} 
  E.~Barausse, V.~Cardoso and G.~Khanna,
  Phys.\ Rev.\ Lett.\  {\bf 105}, 261102 (2010).

\bibitem{Barausse:2011vx} 
  E.~Barausse, V.~Cardoso and G.~Khanna,
  Phys.\ Rev.\ D {\bf 84}, 104006 (2011).

\bibitem{Colleoni:2015afa} 
  M.~Colleoni and L.~Barack,
  Phys.\ Rev.\ D {\bf 91}, 104024 (2015).

\bibitem{Colleoni:2015ena} 
  M.~Colleoni, L.~Barack, A.~G.~Shah and M.~van de Meent,
  Phys.\ Rev.\ D {\bf 92}, no. 8, 084044 (2015).

\bibitem{Hubeny:1998ga} 
  V.~E.~Hubeny,
  Phys.\ Rev.\ D {\bf 59}, 064013 (1999).

\bibitem{Isoyama:2011ea} 
  S.~Isoyama, N.~Sago and T.~Tanaka,
  Phys.\ Rev.\ D {\bf 84}, 124024 (2011).

\bibitem{Aniceto:2015klq} 
  P.~Aniceto, P.~Pani and J.~V.~Rocha,
JHEP {\bf 1605}, 115 (2016).

\bibitem{Hod:2016hqx} 
  S.~Hod,
Class.\ Quant.\ Grav.\  {\bf 33}, no. 3, 037001 (2016).

\bibitem{Horowitz:2016ezu} 
  G.~T.~Horowitz, J.~E.~Santos and B.~Way,
Class.\ Quant.\ Grav.\  {\bf 33}, no. 19, 195007 (2016).

\bibitem{Toth:2015cda} 
  G.~Z.~Toth,
Class.\ Quant.\ Grav.\  {\bf 33}, no. 11, 115012 (2016).
  
\bibitem{Rocha:2011wp} 
  J.~V.~Rocha and V.~Cardoso,
  Phys.\ Rev.\ D {\bf 83}, 104037 (2011).


\bibitem{Gwak:2012hq} 
  B.~Gwak and B.~H.~Lee,
  Class.\ Quant.\ Grav.\  {\bf 29}, 175011 (2012).

\bibitem{Gwak:2015sua} 
  B.~Gwak and B.~H.~Lee,
Phys.\ Lett.\ B {\bf 755}, 324 (2016).

\bibitem{BouhmadiLopez:2010vc} 
  M.~Bouhmadi-Lopez, V.~Cardoso, A.~Nerozzi and J.~V.~Rocha,
  Phys.\ Rev.\ D {\bf 81}, 084051 (2010).

\bibitem{Doukas:2010be} 
  J.~Doukas,
  Phys.\ Rev.\ D {\bf 84}, 064046 (2011).

\bibitem{Lehner:2010pn} 
  L.~Lehner and F.~Pretorius,
  Phys.\ Rev.\ Lett.\  {\bf 105}, 101102 (2010).

\bibitem{Gwak:2011rp} 
  B.~Gwak and B.~H.~Lee,
  Phys.\ Rev.\ D {\bf 84}, 084049 (2011).

\bibitem{Figueras:2015hkb} 
  P.~Figueras, M.~Kunesch and S.~Tunyasuvunakool,
  Phys.\ Rev.\ Lett.\  {\bf 116}, no. 7, 071102 (2016).

\bibitem{Gwak:2015fsa} 
  B.~Gwak and B.~H.~Lee,
JCAP {\bf 1602}, 015 (2016).

\bibitem{Gwak:2016gwj} 
  B.~Gwak,
  Phys.\ Rev.\ D {\bf 95}, no. 12, 124050 (2017).

\bibitem{Hawking:1974sw}
S.~W. Hawking {\em Commun. Math. Phys.} {\bf 43}, 199-220 (1975).

\bibitem{Hawking:1976de}
S.~W. Hawking {\em Phys. Rev.} {\bf D13}, 191-197 (1976).

\bibitem{Bekenstein:1973ur} 
  J.~D.~Bekenstein,
  Phys.\ Rev.\ D {\bf 7}, 2333 (1973).

\bibitem{Bekenstein:1974ax} 
  J.~D.~Bekenstein,
  Phys.\ Rev.\ D {\bf 9}, 3292 (1974).

\bibitem{Myers:1986un} 
  R.~C.~Myers and M.~J.~Perry,
  Annals Phys.\  {\bf 172}, 304 (1986).

\bibitem{Emparan:2003sy}
R.~Emparan and R.~C. Myers {\em JHEP} {\bf 09}, 025 (2003).

\bibitem{Shibata:2009ad}
M.~Shibata and H.~Yoshino {\em Phys. Rev.} {\bf D81}, 021501 (2010).

\bibitem{Dias:2009iu}
O.~J.~C. Dias, P.~Figueras, R.~Monteiro, J.~E. Santos, and R.~Emparan {\em
  Phys. Rev.} {\bf D80}, 111701 (2009).

\bibitem{Dias:2010eu}
O.~J.~C. Dias, P.~Figueras, R.~Monteiro, H.~S. Reall, and J.~E. Santos {\em
  JHEP} {\bf 05}, 076 (2010).

\bibitem{Dias:2010maa}
O.~J.~C. Dias, P.~Figueras, R.~Monteiro, and J.~E. Santos {\em Phys. Rev.} {\bf
  D82}, 104025 (2010).

\bibitem{Durkee:2010qu}
M.~Durkee and H.~S. Reall {\em Class. Quant. Grav.} {\bf 28}, 035011 (2011).

\bibitem{Murata:2012ct}
K.~Murata {\em Class. Quant. Grav.} {\bf 30}, 075002 (2013).

\bibitem{Dias:2014eua}
O.~J.~C. Dias, G.~S. Hartnett, and J.~E. Santos {\em Class. Quant. Grav.} {\bf
  31}, 245011 (2014), no.~24.

\bibitem{Gwak:2014xra}
B.~Gwak and B.-H. Lee {\em Phys. Rev.} {\bf D91}, 064020 (2015), no.~6.

\bibitem{Gwak:2015ysa}
B.~Gwak, B.-H. Lee, and D.~Ro {\em Phys. Lett.} {\bf B761}, 437-443 (2016).

\bibitem{Ahn:2014fwa}
W.-K. Ahn, B.~Gwak, B.-H. Lee, and W.~Lee {\em Eur. Phys. J.} {\bf C75}, 372
  (2015), no.~8.

\bibitem{Hawking:1971tu}
S.~W. Hawking {\em Phys. Rev. Lett.} {\bf 26}, 1344-1346 (1971).

\bibitem{Schiff:1960gi}
L.~I. Schiff {\em Proc. Nat. Acad. Sci.} {\bf 46}, 871 (1960).

\bibitem{Wilkins:1970wap}
D.~Wilkins {\em Annals of Physics} {\bf 61}, 277 - 293 (1970), no.~2.

\bibitem{Mashhoon:1971nm}
B.~Mashhoon {\em J. Math. Phys.} {\bf 12}, 1075-1077 (1971).

\bibitem{Wald:1972sz}
R.~M. Wald {\em Phys. Rev.} {\bf D6}, 406-413 (1972).

\bibitem{Gwak:2016cbq} 
  B.~Gwak and B.~H.~Lee,
JHEP {\bf 1607}, 079 (2016).

\bibitem{Gwak:2016icd} 
  B.~Gwak and D.~Ro,
arXiv:1610.04847 [gr-qc].

\bibitem{Smarr:1976qy}
L.~Smarr, A.~Cadez, B.~S. DeWitt, and K.~Eppley {\em Phys. Rev.} {\bf D14},
  2443-2452 (1976).

\bibitem{Smarr:1977fy}
L.~Smarr {\em Phys. Rev.} {\bf D15}, 2069-2077 (1977).

\bibitem{Smarr:1977uf}
L.~Smarr and J.~W. York, Jr. {\em Phys. Rev.} {\bf D17}, 2529-2551 (1978).

\bibitem{Witek:2010xi}
H.~Witek, M.~Zilhao, L.~Gualtieri, V.~Cardoso, C.~Herdeiro, A.~Nerozzi, and
  U.~Sperhake {\em Phys. Rev.} {\bf D82}, 104014 (2010).

\bibitem{Bantilan:2014sra}
H.~Bantilan and P.~Romatschke {\em Phys. Rev. Lett.} {\bf 114}, 081601 (2015),
  no.~8.

\bibitem{Bednarek:2015dga}
W.~Bednarek and P.~Banasinski {\em Astrophys. J.} {\bf 807}, 168 (2015), no.~2.

\bibitem{Hirotani:2015fxp}
K.~Hirotani and H.-Y. Pu {\em Astrophys. J.} {\bf 818}, 50 (2016), no.~1.

\bibitem{Sperhake:2015siy}
U.~Sperhake, E.~Berti, V.~Cardoso, and F.~Pretorius {\em Phys. Rev.} {\bf D93},
  044012 (2016), no.~4.

\bibitem{Barkett:2015wia}
K.~Barkett et~al. {\em Phys. Rev.} {\bf D93}, 044064 (2016), no.~4.

\bibitem{Hinderer:2016eia}
T.~Hinderer et~al. {\em Phys. Rev. Lett.} {\bf 116}, 181101 (2016), no.~18.

\bibitem{Konoplya:2016pmh}
R.~Konoplya and A.~Zhidenko,
Phys.\ Lett.\ B {\bf 756}, 350 (2016).

\bibitem{Nojiri:2005vv} 
  S.~Nojiri, S.~D.~Odintsov and M.~Sasaki, Phys.\ Rev.\ D {\bf 71}, 123509 (2005).
  
\bibitem{DeWitt:1964}
  C.~DeWitt and B.~DeWitt, Relativity, Groups and Topology, vol.~12, Gordon \& Breach, p.~719 (1964).
  
\bibitem{Sudarsky:2002mk} 
  D.~Sudarsky and J.~A.~Gonzalez, Phys.\ Rev.\ D {\bf 67}, 024038 (2003).

\bibitem{Hendi:2010gq} 
  S.~H.~Hendi, A.~Sheykhi and M.~H.~Dehghani,
  Eur.\ Phys.\ J.\ C {\bf 70}, 703 (2010).

\bibitem{Hendi:2015xya} 
  S.~H.~Hendi, A.~Sheykhi, S.~Panahiyan and B.~Eslam Panah,
  Phys.\ Rev.\ D {\bf 92}, no. 6, 064028 (2015).


\bibitem{Krolak:1987ofj} 
  A.~Krolak and B.~F.~Schutz,
  Gen.\ Rel.\ Grav.\  {\bf 19}, 1163 (1987).

\bibitem{Ade:2015xua} 
  P.~A.~R.~Ade {\it et al.} [Planck Collaboration],
  Astron.\ Astrophys.\  {\bf 594}, A13 (2016).


\end{thebibliography}
\end{document}